\documentclass[12pt,a4paper]{article}

\usepackage{amsmath,tikz,a4wide}
\usepackage{amsfonts,epsfig,color,latexsym}
\usepackage{amssymb}
\usepackage[english, USenglish]{babel}
\usepackage{epsfig, psfrag}
\usepackage{a4wide}
\usepackage{rotating}

\usepackage{hyperref}

\usetikzlibrary{positioning,arrows}
\usetikzlibrary{calc,backgrounds}
\usetikzlibrary{decorations.pathmorphing}
\usetikzlibrary{decorations.markings}

\csname @addtoreset\endcsname{equation}{section}

\makeatletter
\def\timenow{\@tempcnta\time
  \@tempcntb\@tempcnta
  \divide\@tempcntb60
  \ifnum10>\@tempcntb0\fi\number\@tempcntb
  \multiply\@tempcntb60
  \advance\@tempcnta-\@tempcntb
  :\ifnum10>\@tempcnta0\fi\number\@tempcnta}
\makeatother
\def\oonoo#1#2#3{\vbox{\ialign{##\crcr
	\hfil\hfil\hfil{$#3{#1}$}\hfil\crcr\noalign{\kern1pt\nointerlineskip}
	$#3{#2}$\crcr}}}
\def\oon#1#2{\mathchoice{\oonoo{#1}{#2}{\displaystyle}}
	{\oonoo{#1}{#2}{\textstyle}}{\oonoo{#1}{#2}{\scriptstyle}}
	{\oonoo{#1}{#2}{\scriptscriptstyle}}}
\def\dt#1{\oon{\hbox{\bf .}}{#1}}  
\def\ddt#1{\oon{\hbox{\bf .\kern-1pt.}}#1}
\def\slap#1#2{\setbox0=\hbox{$#1{#2}$}
	#2\kern-\wd0{\hfuzz=1pt\hbox to\wd0{\hfil$#1{/}$\hfil}}}

\begin{document}
\renewcommand{\thefootnote}{\fnsymbol{footnote}}
\newpage
\pagestyle{empty}
\setcounter{page}{0}

%%%%%%%%%%%%%%%%%%%%%%%%%%%%%%%%%%%%%

\newcommand{\norm}[1]{{\protect\normalsize{#1}}}
\newcommand{\p}[1]{(\ref{#1})}
\newcommand{\half}{{\ts \frac{1}{2}}}
\newcommand \vev [1] {\langle{#1}\rangle}
\newcommand \ket [1] {|{#1}\rangle}
\newcommand \bra [1] {\langle {#1}|}

\newcommand{\cD}{{\cal D}}
\newcommand{\cM}{{\cal M}} 
\newcommand{\cR}{{\cal R}} 
\newcommand{\cS}{{\cal S}} 
\newcommand{\cK}{{\cal K}}
\newcommand{\cL}{{\cal L}} 
\newcommand{\cE}{{\cal E}}
\newcommand{\cF}{{\cal F}}
\newcommand{\cN}{{\cal N}}
\newcommand{\cA}{{\cal A}}
\newcommand{\cB}{{\cal B}}
\newcommand{\cG}{{\cal G}}
\newcommand{\cO}{{\cal O}}
\newcommand{\cY}{{\cal Y}}
\newcommand{\cX}{{\cal X}}
\newcommand{\cT}{{\cal T}}
\newcommand{\cW}{{\cal W}}
\newcommand{\cP}{{\cal P}}
\newcommand{\nt}{\notag\\} 
\newcommand{\pa}{\partial}
\newcommand{\ep}{\epsilon}
\newcommand{\om}{\omega}
\newcommand{\bep}{\bar\epsilon}
\renewcommand{\a}{\alpha}
\renewcommand{\b}{\beta}
\newcommand{\g}{\gamma}
\newcommand{\s}{\sigma}
\newcommand{\la}{\lambda}
\newcommand{\tl}{\tilde\lambda}
\newcommand{\tm}{\tilde\mu}
\newcommand{\da}{{\dot\alpha}}
\newcommand{\db}{{\dot\beta}}
\newcommand{\dg}{{\dot\gamma}}
\newcommand{\dd}{{\dot\delta}}
\newcommand{\q}{\theta}
\newcommand{\bq}{\bar\theta}
\newcommand{\bQ}{\bar Q}
\newcommand{\tx}{\tilde{x}}
\newcommand{\tr}{\mbox{tr}}
\newcommand{\+}{{\dt+}}
\renewcommand{\-}{{\dt-}}
\newcommand{\ti}{{\textup{i}}}

\vspace{20mm}

%\bigskip  \timenow\ \today   \centerline{File name: {\it draft-v4}}
\bigskip
\bigskip

%\newpage
\setcounter{page}{1}\setcounter{footnote}{0}
\pagestyle{plain}
\renewcommand{\thefootnote}{\arabic{footnote}}

%\newpage\setcounter{page}1

\tikzset{
particle/.style={thick,draw=blue, postaction={decorate},
    decoration={markings,mark=at position .5 with {\arrow[blue]{triangle 45}}}},
gluon/.style={decorate, draw=black,
    decoration={coil,aspect=0}},
photon/.style={decorate, decoration={snake}},
dot/.style={circle,fill=black,inner sep=0pt,minimum size=.2cm},
dotg/.style={circle,fill=black!25,inner sep=0pt,minimum size=.2cm},
dotw/.style={circle,draw=black,inner sep =0cm,minimum size=.2cm} }

\thispagestyle{empty}

\null\vskip-43pt \hfill
\begin{minipage}[t]{50mm}
CERN-TH-2016-253 \\
DCPT-16/57 \\
IPhT--T16/166 \\
LAPTH-072/16 \\
MITP/16-141
\end{minipage}

\vskip2.2truecm
\begin{center}
\vskip 0.2truecm

 {\Large\bf
%\titleline
Wilson Loop Form Factors: A New Duality}
\vskip 0.5truecm
%\vfill

\vskip 1truecm
%\vfill
{\bf  Dmitry Chicherin$^{a}$, Paul Heslop$^{b}$,  \\[2mm] Gregory P. Korchemsky$^{c}$,  
Emery Sokatchev$^{d,e}$ \\
}

\vskip 0.4truecm

% \addresses
$^{a}$
{\it PRISMA Cluster of Excellence, Johannes Gutenberg University,
55099 Mainz, Germany\\
  \vskip .2truecm
$^{b}$   Mathematics Department, Durham University,
Science Laboratories,
\\South Rd, Durham DH1 3LE,
United Kingdom \\
 \vskip .2truecm
$^{c}$ Institut de Physique Th\'eorique\,\footnote{Unit\'e de Recherche Associ\'ee au CNRS UMR 3681},
CEA Saclay, %\\
91191 Gif-sur-Yvette Cedex, France\\
\vskip .2truecm
 $^{d}$ LAPTh\,\footnote[2]{Laboratoire d'Annecy-le-Vieux de Physique Th\'{e}orique, UMR 5108},   Universit\'{e} de Savoie, CNRS,
B.P. 110,  F-74941 Annecy-le-Vieux, France\\
\vskip .2truecm $^{e}$ Theoretical Physics Department, CERN, CH -1211, Geneva 23, Switzerland 
                       } \\
\end{center}

\vskip 5mm

\centerline{\bf Abstract} % \normalsize
\bigskip
\noindent
We find a new duality  for form factors of lightlike Wilson loops in planar $\mathcal N=4$ super-Yang-Mills theory. The duality maps a form factor involving an $n$-sided lightlike polygonal super-Wilson loop together with $m$ external on-shell states, to the same type of object  but with the edges of the Wilson loop and the external states swapping roles.  This relation can essentially be seen graphically in Lorentz harmonic chiral (LHC) superspace 
where it is equivalent to planar graph duality. However there are some crucial subtleties  with the cancellation of spurious poles due to the gauge fixing. They are resolved by finding the correct formulation of the Wilson loop and by careful analytic continuation from Minkowski to Euclidean space. We illustrate all of these subtleties explicitly in the simplest non-trivial  NMHV-like  case.

\newpage

\thispagestyle{empty}

{\small \tableofcontents}

\newpage
\setcounter{page}{1}\setcounter{footnote}{0}

\section{Introduction}
\label{sec:introduction}

The natural gauge invariant objects in any gauge theory include scattering amplitudes, Wilson loops, correlation functions and form factors of local operators. In the past years numerous studies have revealed interesting duality relations between the first three objects in planar $\mathcal N=4$ SYM theory. The {simplest MHV gluon} scattering amplitude $\mathrm{A}_n(p_1,\ldots, p_n)$  has been shown \cite{Alday:2007hr,Drummond:2007aua,Brandhuber:2007yx} to be dual to a Wilson loop $\cW_n(x_1,\ldots,x_n)$ defined on a lightlike contour, 
\begin{align}\label{1.1}
\mathrm{A}_n(p_1,\ldots, p_n) = \cW_n(x_1,\ldots,x_n)\,, 
\end{align}
upon the identification of the separation between the cusp points $x_i$ of the contour with the particle momenta $p_i$ in Minkowski space, $x_i - x_{i+1} = p_i $
for $i=1,\ldots, n$ and $x_{n+1}\equiv x_1$.
This duality has a natural supersymmetric extension \cite{CaronHuot:2010ek,Mason:2010yk,Belitsky:2011zm} where the super-lightlike contour  
{is built out of} the on-shell supermomenta  of the scattered particles. The correlation functions $G_n=\vev{O(x_1)\ldots O(x_n)}$ of local gauge invariant operators $O(x)$ are dual to the Wilson loops (and hence to the amplitudes) in the lightlike limit \cite{Alday:2010zy,Eden:2010zz}, $
\lim_{x^2_{i,i+1}\to0} x^2_{12} \ldots x^2_{n1}\ G_n = \cW_n $. 
This duality has a supersymmetric generalisation as well~\cite{Eden:2011yp,Eden:2011ku,Adamo:2011dq}. 

The fourth object is the form factor $\bra{0}O(x)\ket{k_1,\ldots,k_m}$ of a local operator $O(x)$ with an asymptotic $m-$particle state of on-shell momenta $k_j^2=0$ for $j=1,\ldots,m$. It  is  a hybrid between correlation functions and scattering amplitudes because it lives simultaneously in coordinate and momentum spaces. Such form factors (and their supersymmetric extensions in $\cN=4$ SYM) have been actively studied in the recent years \cite{Brandhuber:2010ad,Brandhuber:2011tv,Engelund:2012re,Koster:2016loo,Koster:2016fna,Bork:2016hst,Bork:2016xfn}. It is interesting  to know if there are possible duality relations for them as well. This question has been addressed in \cite{Derkachov:2013bda} but for a more complicated object, the matrix element of a lightlike bosonic Wilson loop stretched between local operators 
along a single light-cone direction, with an on-shell state. It has been shown that this object is dual to itself upon swapping the coordinate and momentum data. It has  also been conjectured there that the new duality may extend to a larger class of objects, namely the form factor $W_{n,m} =\bra{0}\cW_n(x_1,\ldots,x_n)\ket{k_1,\ldots,k_m}$ of an $n-$gon lightlike (supersymmetric)  Wilson loop with an $m-$particle state. Schematically, the suggested duality takes the form 
\begin{align}\label{1.4}
W_{n,m}(\{x\}|\{k\}) = W_{m,n}(\{y\}|\{p\}) \,,
\end{align}
where the kinematical data on both sides are related like in \p{1.1},
\begin{align}\label{1.5}
x_i - x_{i+1} = p_i \, ,\qquad   \qquad y_j - y_{j+1} = k_j  \, , 
\end{align}
for $i=1,\ldots,n$ and $j=1,\ldots,m$   provided that the total momenta of the particles vanish, $\sum_{i=1}^n p_i= \sum_{j=1}^m k_j=0$.
This conjecture has been successfully tested in \cite{Simon} in the simplest case of a Wilson loop with a state of helicity $(+1)$ gluons and in the Born approximation. 

Building upon the observations in \cite{Derkachov:2013bda} and \cite{Simon}, in this paper we study the general case of the form factor for a lightlike supersymmetric Wilson loop and  we argue that it has a remarkable duality property in planar $\mathcal N=4$ SYM. It extends the bosonic relation \p{1.4}  and the identification of coordinates with momenta \p{1.5} to their supersymmetric analogs. The super-Wilson loop form factors are considered in the planar limit and in the lowest-order perturbative approximation (Born level). The introduction of Grassmann variables ($\q_i$ on the Wilson loop contour and $\eta_j$ for the on-shell states) allows us to probe the duality for more complicated configurations of particle  helicities. By analogy with the amplitudes, we call the contributions at the lowest level in the Grassmann expansion MHV-like, at the next level NMHV-like, etc. At MHV level we confirm the result of  \cite{Simon}. The NMHV level is much more complicated, the form factor being a non-trivial rational function of the kinematical data. Yet, we show that the duality still works, in a rather simple and suggestive way, by just matching planar Feynman diagrams. This allows us to  argue that it should hold for the complete supersymmetric object (at all Grassmann levels) and also beyond the Born approximation.  

The key to understanding the duality is the appropriate superspace formulation of the Wilson loop and its form factor.    In the conventional approach the chiral supersymmetric Wilson loop \cite{CaronHuot:2010ek,Mason:2010yk,Belitsky:2011zm}  is formulated in terms of constrained on-shell super-connections \cite{Harnad:1985bc,Ooguri:2000ps}, which makes the Feynman diagram technique  highly inefficient. In this paper we prefer to use the Lorentz harmonic chiral  (LHC) superspace  approach  \cite{Chicherin:2016fac}. It provides an off-shell formulation of the chiral $\cN=4$ SYM theory in terms of {unconstrained} prepotentials, best suited for supersymmetric quantisation. LHC superspace is an alternative  to the twistor formulation \cite{Mason:2005zm,Boels:2006ir}, closer in spirit to traditional field theory (see also \cite{Rosly:1996vr}).  The main idea is to consider the interacting theory as a perturbation of the self-dual sector. The twistor formulation has been successfully used to justify the so-called MHV rules for the computation of the amplitude \cite{Bullimore:2010pj}, to prove the duality between  supersymmetric Wilson loops and amplitudes \cite{Mason:2010yk}, to compute off-shell correlation functions of the $\cN=4$ stress-tensor multiplet \cite{Chicherin:2014uca}. More recently, the LHC formalism was applied to finding the non-chiral completion of the correlators \cite{Chicherin:2016fbj} and to  the calculation of form factors  of local operators \cite{Chicherin:2016qsf}. In this paper, after explaining the kinematical setup in Section~2,  we formulate the lightlike Wilson loop in LHC superspace in Section~\ref{s2} and apply the Feynman rules of \cite{Chicherin:2016qsf} to the computation of its form factors in Section~4. We find an important additional contribution to the Wilson loop,  compared  to the twistor formulation \cite{Mason:2010yk}. It is needed to make the Wilson loop gauge invariant. 

The duality essentially works on a graph-to-graph basis. More precisely, we find two types of Feynman graphs corresponding to two different helicity configurations at NMHV-like level. These graphs are dual to each other after identifying the kinematical data as in \p{1.5} and redrawing the graph following a simple rule. In addition to these graphs there are sets of graphs whose role is to restore gauge invariance. We use a light-cone gauge whose parameter is the so-called reference spinor. A known problem of such gauges is the presence of spurious poles. Their elimination in the Feynman graphs (and hence the restoration of gauge invariance) is a somewhat subtle  procedure which we describe in detail in Section~\ref{sec:nmhv}. 

We end the paper with several appendices. In Appendix \ref{apA} we explain how to obtain the LHC formulation of the lightlike Wilson loop starting from the standard one with constrained super-connections. In Appendix \ref{apB} we summarise the Feynman rules in the light-cone gauge. In Appendix \ref{aC} we derive some  Fourier transforms that we need for establishing the duality. In Appendix \ref{sec:boundary-cases} we explain the mechanism of spurious pole cancellation in the boundary cases. 

\section{Definitions  and summary of the results}

\subsection{Generalised form factors of Wilson loops}

In this paper, we study a new object -- the generalised form factor of the lightlike Wilson loop.
In $\mathcal N=4$ SYM with gauge group $SU(N)$ it is defined as the matrix element of a lightlike $n-$gon supersymmetric 
Wilson loop 
$\mathcal W_n$ with the on-shell $m-$particle state $\ket{1^{a_1}\dots m^{a_m}}$:
\begin{align}\label{W-def0}
\vev{0|\mathcal W_n|1^{a_1}\dots m^{a_m}} ={1\over N}
\vev{0|\tr\, P \exp\left[i\oint_{\mathcal C_n} \Big(dx^\mu \mathcal A_\mu(x,\theta) + d\theta^{\alpha A} {\mathcal A}_{\alpha A}(x,\theta) \Big)\right]|1^{a_1}\dots m^{a_m}} \,,
\end{align}
where the integration goes over a closed contour $\mathcal C_n$ formed by $n$ straight lightlike segments connecting
the superspace points $(x_i,\theta_i)$. The bosonic and fermionic gauge connections, $\mathcal A_\mu$ and ${\mathcal A}_{\alpha A}$, have expansions in powers of $\theta$'s with  coefficients given in terms of  the gluon, gaugino and scalar fields.
Their explicit expressions are shown below in (\ref{con}). 

In the planar limit, the form factor can be decomposed in the standard 
manner over the basis of single traces,
\begin{align}\label{W-def}
\vev{0|\mathcal W_n|1^{a_1}\dots m^{a_m}} = \sum_{\sigma\in S_m/Z_m} \tr (T^{a_{\sigma_1}} \dots T^{a_{\sigma_m}})  F_{n,m}(\sigma_1,\dots,
\sigma_m)\,,
\end{align}
where the sum runs over all permutations of the external particles $\sigma_1,\dots,\sigma_m$ modulo cyclic shifts. 
The matrix element (\ref{W-def}) is a natural generalisation of lightlike Wilson loops $\vev{0|\mathcal W_n|0}$ and scattering amplitudes $\mathrm{A}(1^{a_1}\dots m^{a_m})$. 
In fact, it gets a disconnected contribution given by their product. In what follows we discard it and consider only the connected contribution to
(\ref{W-def}).  

The color-ordered form factors $F_{n,m}$ depend on two sets of variables. The first set consists of $n$ coordinates
in Minkowski space-time and their odd superpartners  $(x_i^{\dot\alpha\alpha},\theta_i^{\alpha A})$ specifying the position 
of the vertices of a lightlike $n-$gon,\footnote{We use two-component spinor notation for vectors, e.g., $x^{\da\a} =(\sigma_\mu)^{\da\a} x^\mu$. The Lorentz and R symmetry indices take values $\a=1,2$, $\da=1,2$ and  $A=1,2,3,4$, respectively. }
\begin{align}\label{x}
(x_i-x_{i+1})^2 = 0 \,,\qquad \qquad   (x_i-x_{i+1})^{\dot\alpha\alpha}\,(\theta_{i,\alpha}^A-\theta_{i+1,\alpha}^A) =0
\end{align}
  for $i=1,\dots,n$, with the cyclicity conditions $x_{n+1}=x_1$ and $\q_{n+1}=\q_1$. Here the first relation means that the Wilson loop is built from lightlike segments and the second
relation is its superpartner.  

The second
set of variables consists of the on-shell momenta of $m$ particles $(k^{\dot\alpha\alpha}_j,\eta_{jA})$
\begin{align}
k_j^{\dot\alpha\alpha} = \tilde k_j^{\dot\alpha}  k_j^{\alpha} \equiv |k_j] \bra{k_j}
\end{align}
with $k_j^2=0$ and $j=1,\dots,m$. Like the scattering amplitudes, the expansion of the on-shell state in powers of  $\eta_{jA}$ corresponds to particles with different helicity (gluons, gaugini and scalars). Each particle superstate carries one unit 
of helicity. It is then convenient to introduce the helicity-free function $W_{n,m}$ {multiplying (\ref{W-def}) by the so-called
Parke-Taylor factor}
\begin{align}\label{Anm}
W_{n,m} = { \vev{k_1k_2} \vev{k_2k_3} \dots \vev{k_m k_1}}\ F_{n,m}(1,\dots,m)\ ,
\end{align} 
where $\vev{k_i k_j} = k^\a_i \ep_{\a\b} k^\b_j$. 
The scalar function $W_{n,m}$ defined in this way depends on the two sets of variables introduced above,
\begin{align}\label{W}
W_{n,m} = W_{n,m}(\{x,\theta\}; \{k,\eta\} )\,.
\end{align}
As follows from the definition (\ref{W-def}), this function is invariant under cyclic shifts
of the coordinates and momenta.

\subsection{Dual variables}

To elucidate the interesting properties of $W_{n,m}$ we introduce the so-called dual superspace variables \cite{Drummond:2008vq}. The coordinates of the
Wilson loop $(x_i,\theta_i^{A})$ have the dual momenta $(p_i,\omega_i^A)$ defined as
\begin{align}\label{p}
x_i - x_{i+1} = p_i\,,\qquad\qquad \ket{\theta_i^A} - \ket{\theta_{i+1}^A} = \ket{p_i} \, \omega_i^A\,,
\end{align}
where we do not display the Lorentz indices for simplicity. It follows from (\ref{x}) that $p_i$ are lightlike vectors,
$p_i^2=0$, satisfying the condition $\sum_{i=1}^n p_i=0$. Similarly, the odd variables $\omega_i^A$ satisfy 
the relation $\sum_{i=1}^n  \ket{p_i} \omega_i^A=0$ and solve the second condition in (\ref{x}). Note that 
the properties of $(p_i,\omega_i^A)$ (with $i=1,\dots,n$) match those of the supermomenta of the on-shell states in the scattering amplitude 
$\mathrm{A}_n$. This observation was crucial  in establishing the duality between the lightlike Wilson loop $\cW_n$
and the scattering amplitude $\mathrm{A}_n$.

For the set of on-shell momenta $(k_j,\eta_{jA})$, the dual coordinates are defined as
\begin{align}\label{k}
k_j = y_j - y_{j+1}\,,\qquad\qquad \ket{k_j} \, \eta_{jA} = \ket{\psi_{j,A}} - \ket{\psi_{j+1,A}}\,.
\end{align} 
Here the dual momenta $y_1,\dots, y_{m+1}$ are consecutively lightlike separated, $(y_i-y_{i+1})^2=0$ and their
superpartners satisfy $(y_j - y_{j+1})(\ket{\psi_{j,A}} - \ket{\psi_{j+1,A}})=0$. 
Note the striking similarity between relations (\ref{p}) and (\ref{k}). Namely, these relations can be
mapped into each other by exchanging  coordinates with dual momenta, $(x,\theta)\to (y,\psi)$, and momenta with
dual coordinates, $(k,\eta)\to (p,\omega)$.
  
There is however an important difference between the two sets of dual coordinates. The dual vectors $p_i$ define
the edges of a closed $n-$gon and their sum equals zero. The same is true for the sum of dual odd
coordinates $\ket{p_i} \, \omega_i^A$,
\begin{align}\label{zero}
\sum_{i=1}^n p_i = x_1 - x_{n+1} = 0\,,\qquad\qquad \sum_{i=1}^n \ket{p_i} \, \omega_i^A= \ket{\theta_1^A}-\ket{\theta_{n+1}^A}=0\,,
\end{align}
so that the dual variables satisfy the periodicity conditions $x_{i} = x_{i+n}$ and $\theta_i^A=\theta_{i+n}^A$.
For the dual momenta the analogous relations read
\begin{align}\label{dual-zero}
 \sum_{j=1}^m k_j = y_1 - y_{m+1}=K\,,\qquad\qquad
 \sum_{j=1}^m \ket{k_j} \, \eta_{jA} = \ket{\psi_{1,A}} - \ket{\psi_{m+1,A}} = Q_A\,,
\end{align}
where $K$ and $Q$ are the total momentum and supercharge of the $m$ particles in (\ref{W-def}), respectively. 
In contrast with (\ref{zero}), $K$ and $Q$ can take arbitrary values and there are no reasons to impose the periodicity
conditions $y_{m+1}=y_1$ and $\psi_{m+1,A}=\psi_{1,A}$. Indeed, the function (\ref{W}) is well defined for arbitrary $K$ and $Q$.

\subsection{Duality relation}
 
Setting $K=Q_A=0$ in (\ref{dual-zero}) we restore the symmetry between (\ref{zero}) and (\ref{dual-zero}). This allows us to 
treat the original variables and their dual counterparts on an equal footing. 
In this paper we argue that for $K=Q_A=0$ the symmetry of $W_{n,m}$ is enhanced and yields an interesting duality relation 
for  $W_{n,m}$ that we shall formulate in a moment. 
More precisely, we can use the dual variables to define,
following  (\ref{W-def}), the matrix element of the lightlike Wilson loop $\vev{0|\cW_m|1^{a_1}\dots n^{a_n}}$. Here the Wilson loop
is evaluated along a closed lightlike $m-$gon with vertices located at $(y_j,\psi_j)$ and the on-shell state consists of $n$ particles with
supermomenta  $(p_i, w_i^A)$. This matrix element has the same general form (\ref{W-def}) and (\ref{Anm}), with the corresponding
scalar function $W_{m,n}$ given by 
\begin{align}\label{W-dual}
W_{m,n} = W_{m,n}(\{y,\psi\}; \{p,\omega\} )\,.
\end{align}
Applying relations (\ref{p}) and (\ref{k}) we can express it in terms of the original variables $\{x_i,\theta_i\}$ and $\{k_j,\eta_j\}$. 

\begin{figure}[t] 
\hspace*{-3mm}
\includegraphics[width=1.05\textwidth]{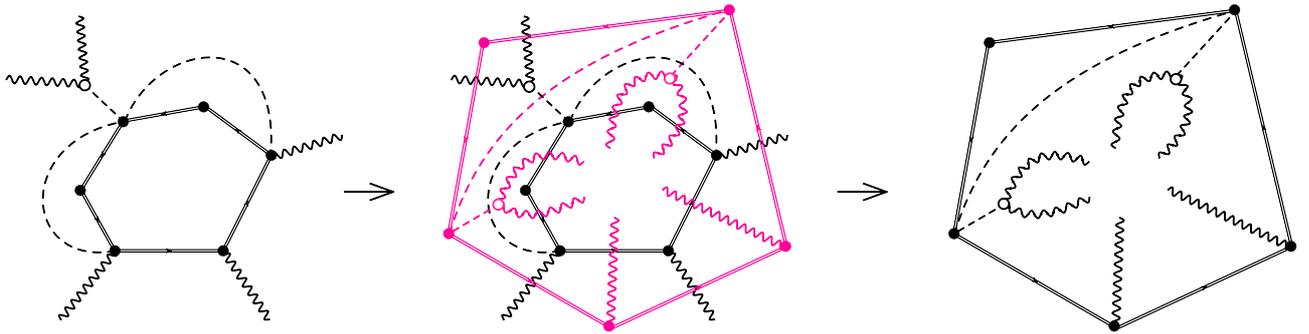}
\caption{Diagrammatic representation of the duality relation (\ref{duality1}). The Wilson loop on the left is built out of lightlike vectors $p_1,\dots, p_n$, the wavy lines denote on-shell particles with momenta $k_1,\dots,k_m$ and the dash lines stand for free propagators. Black and white dots denote effective vertices. The dual Wilson loop form factor on the right  has the lightlike vectors and momenta exchanged. The middle figure explains the duality by superimposing the two graphs. }
\label{fig:NMHV}
\end{figure}

The duality relation that we propose states that the functions   (\ref{W}) and \p{W-dual}   coincide in planar $\mathcal N=4$ SYM,
\begin{align}\label{duality}
W_{n,m}(\{x,\theta\}; \{k,\eta\} )=W_{m,n}(\{y,\psi\}; \{p,\omega\}) \,.
\end{align}
Using the definition of the dual variables we can rewrite the   duality relation in other equivalent forms, e.g.
\begin{align}\label{duality1}
W_{n,m}(\{p,\omega\}; \{k,\eta\} )=W_{m,n}(\{k,\eta\}; \{p,\omega\}) \,.
\end{align}
This relation is represented diagrammatically  in Figure~\ref{fig:NMHV}.  
 
The duality relation (\ref{duality}) should hold for any values of $n$ and $m$. 
As a simple illustration, we examine it for the lowest values of $n$ and $m$. 
In the special cases $m=0$ (or $n=0$) we recover the well-known duality between the $n-$point superamplitude and the $n-$point super-Wilson loop. Since the $n-$gon
Wilson loop is well defined for $n\ge 2$, we start with $n=2,3$. In this case, the cusp points $x_i$ satisfying (\ref{x}) have to 
lie on the same light-ray in Minkowski space-time. Then, the integration contour   of the Wilson loop
collapses  to a  backtracking path leading to $W_{2}=W_{3}=1$. As a consequence, the matrix element on the left-hand 
side of (\ref{W-def}) only receives disconnected contributions yielding the vanishing of $W_{n,m}(\{x,\theta\}; \{k,\eta\})$
for $n=2,3$. The duality relation (\ref{duality}) implies that the same should be true for $W_{m,n}(\{y,\psi\}; \{p,\omega\})$ 
for $n=2,3$. Indeed, the corresponding matrix element (\ref{W-def}) involves an on-shell state with  (real valued) lightlike momenta $k_i$ that
are necessarily aligned due to $\sum_i k_i=0$. In this case $\vev{k_i k_j}=0$ and it follows from (\ref{Anm}) that $W_{m,n}$ vanishes, in agreement with (\ref{W-dual}).

\subsection{Duality relation at MHV level}  
 
Let us now consider the duality relation for $n,m\ge 4$. In this case both sides of (\ref{duality}) are different from zero and are given
by nontrivial functions of the kinematical variables and of the 't Hooft coupling constant. In what follows we shall restrict our 
consideration to the lowest order in the coupling (Born level). Expanding both sides of (\ref{duality}) in the Grassmann variables, 
we can get relations between the different components. By analogy with the scattering amplitudes, we shall refer to the   terms of
the expansion as MHV, NMHV, etc. Notice that since $W_{n,m}(\{x,\theta\}; \{k,\eta\} )$ depends on two sets of
Grassmann variables $\theta_i$ and $\eta_{j}$, we will have to deal with a double expansion of the form N${}^{\kappa}$MHV $\times$N${}^\sigma$MHV.

The lowest term of the expansion,
MHV$\times$MHV, corresponds to (\ref{duality}) with all Grassmann variables put to zero on both sides of the relation. 
Namely, for $\theta_i=0$ the super Wilson loop $\mathcal W_n$ reduces to the bosonic lightlike Wilson loop and for $\eta_j=0$ the on-shell
state in (\ref{W-def}) reduces to a gluon state of helicity $(+1)$. In this way,  from (\ref{W-def0}) and
(\ref{W-def}) we obtain
\begin{align}\label{A-E0}
F_{n,m}^{\rm MHV\times MHV}(x,k) = {1\over N}\vev{0| \tr\big(E_{1n} \dots E_{32} E_{21} \big)|k_1^+ \dots k_m^{+}}\,,
\end{align}
where $E_{i+1,i}$ denotes a bosonic Wilson line in the fundamental of $SU(N)$
evaluated along the lightlike segment $[x_i,x_{i+1}]$  
\begin{align}
E_{i+1,i} =  P \exp\left(-i  \int_0^1 dt \, p_i \cdot A(x_i - p_i t) \right)\,,
\end{align}
with $p_i=x_i -x_{i+1}$.
Notice that the ordering of the $E-$factors inside the trace in (\ref{A-E0}) is opposite to that of the gluons in the on-shell state.

\begin{figure}[t] 
 \psfrag{x1}[cc][cc]{$x_1$}\psfrag{x2}[cc][cc]{$x_2$}\psfrag{x3}[cc][cc]{$x_3$}\psfrag{x4}[cc][cc]{$x_4$}
 \psfrag{x5}[cc][cc]{$x_5$}\psfrag{x6}[cc][cc]{$x_6$}
  \psfrag{y1}[cc][cc]{$y_1$}\psfrag{y2}[cc][cc]{$y_2$}\psfrag{y3}[cc][cc]{$y_3$}\psfrag{y4}[cc][cc]{$y_4$}
 \psfrag{y5}[cc][cc]{$y_5$} 
 \psfrag{p1}[cc][cc]{$p_1$}\psfrag{p2}[cc][cc]{$p_2$}\psfrag{p3}[cc][cc]{$p_3$}\psfrag{p4}[cc][cc]{$p_4$}
 \psfrag{p5}[cc][cc]{$p_5$}\psfrag{p6}[cc][cc]{$p_6$}
 \psfrag{k1}[cc][cc]{$k_1$}\psfrag{k2}[cc][cc]{$k_2$}\psfrag{k3}[cc][cc]{$k_3$}\psfrag{k4}[cc][cc]{$k_4$}
 \psfrag{k5}[cc][cc]{$k_5$} 
\includegraphics[width=\textwidth]{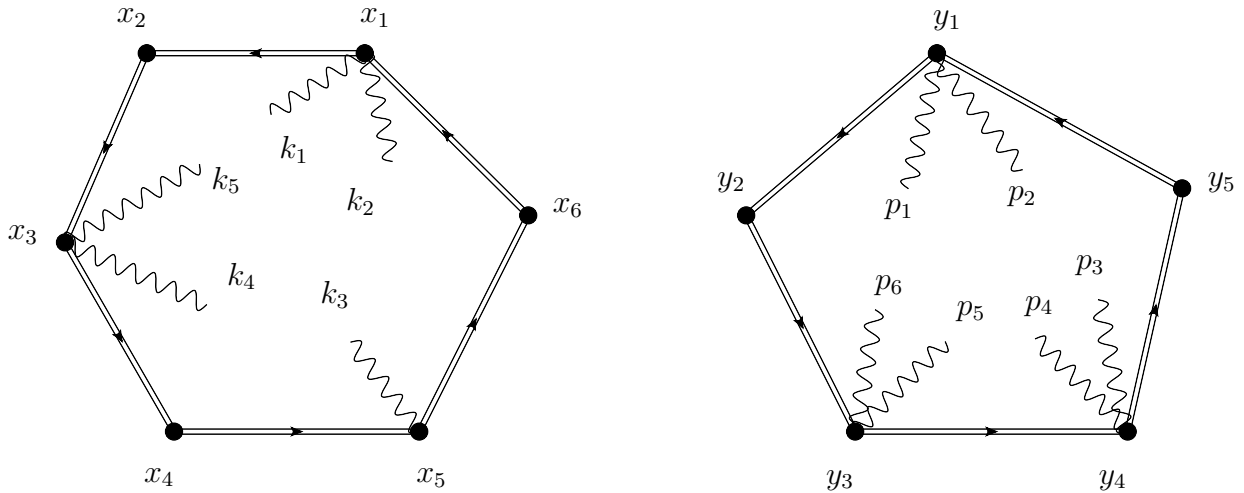}
\caption{Diagrammatic representation of the duality relation (\ref{MHV-dual}) for $n=6$ and $m=5$. Notice that the polygon vertices and the gluons are ordered in opposite directions. Black blobs with outgoing gluons denote wedge form factors (\ref{wedge}).
The lightlike edges of the Wilson loops are mapped to the momenta of the on-shell gluons, $k_i =y_i-y_{i+1}$ and $p_j=x_j-x_{j+1}$.
}
\label{fig:MHV}
\end{figure}

In the Born approximation, $\mathrm{A}_{n,m}^{\rm MHV\times MHV}$ is given by the sum of tree  Feynman diagrams in which the 
on-shell gluons are attached to the lightlike $n-$gon contour either directly or through $3-$ and $4-$gluon interaction vertices. 
The calculation can be simplified by introducing the notion of a ``wedge", i.e. a  cusped Wilson
line built from two semi-infinite rays running along the lightlike vectors $-p_1$ and $p_2$ and joining at point $x$:
\begin{align}
W_{p_2, p_1}(x) = P \left[\exp\left(i\int^\infty_0 dt \, p_2\cdot A(x+p_2 t)\right) \exp\left(- i\int_{-\infty}^0 dt \, p_1\cdot A(x-p_1 t)\right)\right] \,.
\end{align}
In the product $W_{p_3, p_2}(x_3)W_{p_2, p_1}(x_2)$ with $p_2=x_2-x_3$, it is easy to see that the two semi-infinite rays
running along $p_2$ partially cancel against each other giving rise to $E_{32}$. In this way, we can rewrite (\ref{A-E0}) as
 \begin{align}\label{A-E}
F_{n,m}^{\rm MHV\times MHV}(x,k) = {1\over N}\vev{0| \tr\big[W_{p_n, p_{n-1}}(x_n) \dots W_{p_{2}, p_1}(x_2)  W_{p_1, p_n}(x_1) \big]|k_1^+ \dots k_m^{+}}\,.
\end{align}
The advantage of this representation is that, in the Born approximation, the on-shell gluons can be emitted by one of the
$W-$factors thus allowing us to express the matrix element on the right-hand side of (\ref{A-E}) as the sum over all possible
attachments of $m$ gluons to $n$ wedges 
\begin{align}\notag\label{A-factor}
F_{n,m}^{\rm MHV\times MHV} {}& =     
 \ \sum_{  \ell_1 < \dots < \ell_s }  \ \sum_{1\le i_s < \dots < i_1 \le n}
 \vev{0|W_{p_{i_1}, p_{i_1-1}}(x_{i_1})|k_{\ell_1}^+\dots k_{\ell_2-1}^+ }
\\[2mm]
{}& \times
\vev{0|W_{p_{i_2}, p_{i_2-1}}(x_{i_2})|k_{\ell_2}^+\dots k_{\ell_3-1}^+ } \dots
\vev{0|W_{p_{i_s}, p_{{i_s}-1}}(x_{i_s})|k_{\ell_{s}}^+\dots k_{\ell_1-1}^+ } \,.
\end{align}
Here the first sum goes over all possible partitions 
of $m$ gluons over $s$  clusters (with $s\le n$) and the second sum runs over 
all possible wedges $x_{i_1},\dots,x_{i_s}$ to which these clusters are attached. The difference in the ordering of indices $\ell_k$
and $i_k$ in (\ref{A-factor})  is due to the opposite ordering of the $E-$factors and gluons in (\ref{A-E0}).
Relation (\ref{A-factor}) is represented
diagrammatically   in Figure~\ref{fig:MHV}.

Relation (\ref{A-factor}) involves the so-called wedge form factor $\vev{0|W_{p_2, p_1}(x)|k_1^+\dots k_\ell^+}$. Since the on-shell
state contains only gluons of the same helicity, its calculation in the Born approximation can be performed
in the self-dual sector of Yang-Mills theory \cite{Rosly:1996vr,Simon}  
\begin{align}\label{wedge}
 \vev{0|W_{p_2, p_1}(x)|k_1^+\dots k_\ell^+ } = F(p_2,k_1,\dots,k_\ell, p_1)  \, e^{ix(k_1+\ldots+k_\ell)}\,.
\end{align}
Here the dependence on $x$ is fixed by Poincar\'e symmetry and the order of the arguments of the $F-$function matches the color
ordering of the gluons. Its explicit expression reads (see Section~\ref{sec:mhv-example} for more details)
\begin{align}\label{F-fun}
F(p_2,k_1,\dots,k_\ell, p_1) = {\vev{p_2p_1}\over \vev{p_2k_1} \vev{k_1 k_2} \dots \vev{k_\ell p_1}}\,.
\end{align}
Substituting (\ref{wedge}) and (\ref{F-fun}) in (\ref{A-factor}) and matching the result with (\ref{Anm}) we find
\begin{align}\notag\label{MHV-sum}
W_{n,m}^{\rm MHV\times MHV}(x,k) {}&=  \sum   e^{ix_{i_1} y_{\ell_1\ell_2}+ i x_{i_2}y_{\ell_2,\ell_3} + \ldots + i x_{i_s} y_{\ell_{s},\ell_1}}
\\
{}& \times 
{\vev{k_{\ell_1-1} k_{\ell_1}} \vev{ p_{i_1}p_{i_1-1}}\vev{k_{\ell_2-1} k_{\ell_2}}\vev{ p_{i_2}p_{i_2-1}}  \dots
\vev{k_{\ell_s-1} k_{\ell_s}}\vev{p_{i_s}p_{i_s-1} }
 \over \vev{k_{\ell_1}p_{i_1}} \vev{p_{i_1-1}k_{\ell_2-1}} \vev{k_{\ell_2} p_{i_2}}\vev{p_{i_2-1}k_{\ell_3-1}}\dots \vev{k_{\ell_s}p_{i_s}}\vev{p_{i_s-1} k_{\ell_1-1} } } \,,
\end{align}
where the sum covers the same range as in (\ref{A-factor}). Here we used (\ref{k}) to switch to dual momenta  in the exponent, e.g.
$y_{\ell_1\ell_2} = k_{\ell_1} + \ldots + k_{\ell_2-1}$. We recall that for vanishing total momentum $K=\sum_{i=1}^m k_i =0$, the dual
momenta satisfy the periodicity condition $y_{m+1}=y_1$. Using this property, we can rewrite the exponential factor in (\ref{MHV-sum}) in the 
equivalent form
\begin{align}
 e^{ i y_{\ell_s} x_{i_s,i_{s-1}} + \ldots + i y_{\ell_2}x_{i_2 i_1} +  i y_{\ell_1}x_{i_1i_s}  }\,.
\end{align}
We observe that it can be obtained from the original factor by swapping the variables
\begin{align}\label{rel1}
x_{i_1} \leftrightarrow y_{\ell_s}\,,\qquad x_{i_2} \leftrightarrow y_{\ell_{s-1}}\,,\quad \dots\,, \quad
x_{i_s}  \leftrightarrow y_{\ell_1}\,.
\end{align}
Let us now examine the expression in the second line of (\ref{MHV-sum}). It depends on two sets of null vectors $p_i$ and $k_i$
defining the edges of the lightlike Wilson loop and the momenta of the on-shell gluons, respectively. It is straightforward to verify that it is invariant under the swapping of these vectors
\begin{align}\label{rel2}
k_{\ell_1} \leftrightarrow p_{i_s}\,,\qquad k_{\ell_2} \leftrightarrow p_{i_{s-1}}\,,\quad \dots\,,  \quad k_{\ell_s} \leftrightarrow p_{i_1}\,.
\end{align}
Putting together (\ref{rel1}) and (\ref{rel2}), we immediately conclude that the expression on the right-hand side of (\ref{MHV-sum})
is invariant under the exchange of the original variables $(x,k)$ with their dual partners $(y, p)$. This yields the duality relation 
\begin{align}\label{MHV-dual}
W_{n,m}^{\rm MHV\times MHV}(x,k) = W_{m,n}^{\rm MHV\times MHV}(y, p)\,,
\end{align}
in agreement with \cite{Simon}. 
  
\subsection{Duality beyond MHV}

To test the duality relation (\ref{duality}) beyond MHV level, we have to take into account the dependence of the Wilson loop form factor
(\ref{W-def0}) on the Grassmann variables $\theta_i^A$ and $\eta_{j A}$. The dependence on $\eta$ comes from the expansion of
the on-shell super-state in (\ref{W-def0}) over the states of particles (gluons, gaugino and scalars) with different helicity. 

At the same time, the dependence of (\ref{W-def0}) on $\theta$ comes from the expansion of the supersymmetric $n-$gon Wilson loop 
\begin{align}\label{W-super}
\mathcal W_n = {1\over N} \tr \big( \mathcal E_{1n} \dots \mathcal E_{32} \, \mathcal E_{21}\big)
\end{align}
in powers of $\theta_i$ defining the position of vertices of the lightlike $n-$gon in (chiral) superspace. Here the 
supersymmetric Wilson line
$\mathcal E_{i+1,i}$
is evaluated along the straight segment  
connecting the superspace points $(x_i,\theta_i)$ and $(x_{i+1},\theta_{i+1})$ 
\begin{align}\label{127}
\mathcal E_{i+1,i} =  P \exp\left[-i\int_0^1 dt\, \Big(\frac12 x_{i,i+1}^{\dot\alpha\alpha} \mathcal A_{\alpha\dot\alpha}(x(t),\theta(t)) +
\theta_{i,i+1}^{\alpha A} \mathcal A_{\alpha A}(x(t),\theta(t))\Big) \right]\,,
\end{align}
where $x(t) = x_i - x_{i,i+1}\,  t$ and $\theta(t) = \theta_i - \theta_{i,i+1}\, t$.
The super-connections $\cA$ are subject to the defining {\it on-shell} constraints of $\cN=4$ SYM \cite{Sohnius:1978wk}. One way of solving them is to fix the {\it non-supersymmetric} Wess-Zumino gauge and express the components of $\cA$ in terms of the propagating gluon, gaugino and scalar fields \cite{Harnad:1985bc,Ooguri:2000ps}
\begin{align}\notag\label{con}
{}& \mathcal A_{\alpha\dot\alpha} = A_{\alpha\dot\alpha} + i\theta_\alpha^A \bar\psi_{\dot\alpha A} + {i\over 2!} \theta^A_\alpha
\theta^{\beta B} D_{\beta\dot\alpha} \bar\phi_{AB} -{1\over 3!} \epsilon_{ABCD} \theta_\alpha^A\theta^{\beta B} 
\theta^{\gamma C} D_{\beta\dot\alpha} \psi_\gamma^D + \dots
\\
{}& \mathcal A_{\alpha A} = {i\over 2} \bar\phi_{AB} \theta^B_\alpha -{1\over 3!!}\epsilon_{ABCD} \theta^B_\alpha \theta^{\gamma C}
\psi_\gamma^D + {i\over 4!!} \epsilon_{ABCD} \theta^B_\alpha \theta^{\beta C} \theta^{\gamma D} F_{\beta\gamma} + \dots\,,
\end{align}
where the dots denote higher-order terms in $\theta$.
  
Before continuing let us examine the superspace  structure we should expect  this object to have arising from supersymmetry. The chiral supersymmetry  of (\ref{W})   yields the Ward identity 
\begin{align}\label{Ward}
\bigg(\sum_{i=1}^n {\partial\over\partial \theta_i^A} + \sum_{j=1}^m \ket{k_j} \eta_{j,A}\bigg) W_{n,m}(\{x,\theta\}; \{k,\eta\} ) = 0\,.
\end{align} 
The duality relation is expected to hold  if the total particle supercharge vanishes,  $Q_A=\sum_{j=1}^m \ket{k_j} \, \eta_{jA} = 0$. Then (\ref{Ward}) implies that $W_{n,m}$ can be an arbitrary function of $\theta_{ij}^A=\theta_i^A-\theta_j^A$ and $\eta_{k A}$.
In virtue of the R symmetry, these variables must form  $SU(4)$ invariants. The latter are of three 
different kinds: $\epsilon_{ABCD} \theta_{ii'}^A  \theta_{jj'}^B  \theta_{kk'}^C   \theta_{ll'}^D$, 
$\eta^4_{ijkl}=\epsilon^{ABCD} \eta_{i A}\eta_{j B}\eta_{k C}\eta_{l D}$ and $(\theta_{ij}\eta_k)=\theta_{ij}^A\,\eta_{k A}$. 
The dependence on these invariants simplifies further in the Born approximation. 

To compute $W_{n,m}$ in
the Born approximation, we substitute (\ref{W-super}) --  (\ref{con}) into the definition (\ref{W-def0}) and retain 
the contribution at the lowest order in the coupling. Since the dependence on $\theta$'s comes from the expansion of the
bosonic and fermionic connections in (\ref{con}),
the number of contributing diagrams and their complexity increases significantly as compared with the MHV case
described in the previous subsection. Moreover, the use of the Wess-Zumino gauge \p{con} breaks manifest supersymmetry. This makes the conventional approach impractical. 

In this paper we prefer the off-shell formulation of the chiral $\cN=4$ SYM theory in terms of {\it unconstrained} prepotentials in LHC superspace  \cite{Chicherin:2016fac}, better suited for supersymmetric quantisation. In Section~\ref{s2} we formulate the lightlike Wilson loop in LHC superspace and apply the Feynman rules of \cite{Chicherin:2016qsf} to the computation of its form factors.

In this new formulation, $W_{n,m}(\{x_i,\theta_i\}; \{k_j,\eta_j\} )$ 
is given by a sum of contributions having a  similar structure to (\ref{A-factor}), with the important difference that the wedge
form factors are replaced by their supersymmetric generalisations depending on the Grassmann variables $\theta_i^A$ and $\eta_{j A}$.
This leads to the following general expression for $W_{n,m}$,
\begin{align}\label{sum-gen} 
W_{n,m}  {}&=  \sum   e^{ix_{i_1} y_{\ell_1\ell_2}+ i x_{i_2}y_{\ell_2,\ell_3} + \ldots + i x_{i_s} y_{\ell_{s},\ell_1}}
  \times e^{\vev{\theta_{i_1} \psi_{\ell_1\ell_2}}+\vev{\theta_{i_2} \psi_{\ell_2\ell_3}} + \ldots +\vev{\theta_{i_s} \psi_{\ell_s\ell_1}}}\times
\widehat W_{n,m} \,,
\end{align}
which should be compared with (\ref{MHV-sum}). Here we used shorthand notation for $\vev{\theta_{i_1} \psi_{\ell_1\ell_2}} = 
\theta^{\alpha A}_{i_1} (\psi_{\ell_1,\alpha A}-\psi_{\ell_2,\alpha A})$ with the dual $\psi-$variables defined in (\ref{k}). The sum in 
(\ref{sum-gen}) has the same form as in (\ref{A-factor}) and runs over all possible partitions of $m$ super particles over $s$ clusters. 
Notice that the function $\widehat W_{n,m}$ depends on the choice of partition.
The second exponent on the right-hand side of (\ref{sum-gen}) is the supersymmetric completion of the first exponent depending on the bosonic variables. 

Most importantly, as we show below by exploring the structure of the Feynman diagrams, 
the function $\widehat W_{n,m}$ does not depend on the mixed products of
Grassmann variables $(\theta_{ij}\eta_k)$  in the Born approximation.\footnote{This  does not follow from chiral supersymmetry~\eqref {Ward} and it would be interesting to understand the symmetry leading to such a structure.} This allows us to expand $\widehat W_{n,m}$ in powers of the two remaining invariants
leading to the following relation
\begin{align}\label{W-hat}
\widehat W_{n,m} =   W_{n,m}^{(0,0)} +   \left(W_{n,m}^{(1,0)} +   W_{n,m}^{(0,1)}\right)  +    \left(W_{n,m}^{(2,0)} +   W_{n,m}^{(1,1)} +   W_{n,m}^{(0,2)}\right) + \dots\,,
\end{align}  
where $W_{n,m}^{(\kappa ,\sigma )}$ is a homogenous polynomial in $\theta$'s and $\eta$'s of degree $4\kappa$ and $4\sigma$, respectively.
Schematically, $W_{n,m}^{(\kappa ,\sigma )}\sim \theta^{4\kappa}\eta^{4\sigma}$. By analogy with the superamplitude, we refer to the terms on the
right-hand side of (\ref{W-hat}) with $\kappa+\sigma=k$ as N${}^k$MHV-like. 
The lowest term of the expansion, $W_{n,m}^{(0,0)}$, defines the MHV-like contribution $W_{n,m}^{\rm MHV\times MHV}$ 
discussed in the previous subsection. Its explicit expression can be read from (\ref{MHV-sum}).

Substituting  (\ref{sum-gen}) and (\ref{W-hat}) into (\ref{duality}), we can formulate the duality relation in each sector,
\begin{align}\label{duality-gen}
  W_{n,m}^{(\kappa ,\sigma )} (\{x,\theta\}; \{k,\eta\} )=W^{(\sigma,\kappa)} _{m,n}(\{y,\psi\}; \{p,\omega\})\,.
\end{align}
The explicit expressions for $W_{n,m}^{(\kappa ,\sigma )}$ for generic $\kappa$ and $\sigma$ are rather complicated even in the Born approximation. Nevertheless, as we show
below, the duality relation (\ref{duality-gen}) can be verified by matching into each other the diagrams contributing to both sides
of (\ref{duality-gen}).

\section{Lightlike Wilson loop in LHC superspace}\label{s2}

As mentioned in the introduction, the conventional formulation \p{127} of the chiral supersymmetric Wilson loops, making use of constrained super-connections, is not convenient for quantum calculations. The LHC superspace approach, where the dynamical gauge prepotentials are unconstrained, is much more efficient. 
In this section we start by a brief summary of the LHC superspace description of $\cN=4$ SYM. Then we present  the explicit form of the Wilson loop   in LHC superspace, in terms of the two unconstrained gauge prepotentials (the detailed derivation is shown in Appendix \ref{apA}). Our formulation is similar to the twistor one of Mason and Skinner in \cite{Mason:2010yk} but differs from it on an essential point.

\subsection{$\cN=4$ super-Yang-Mills in LHC superspace}

Here we recall some basic facts about  $\cN=4$ SYM in LHC superspace  (for details see \cite{Chicherin:2016fac}).
The theory is formulated in terms of two dynamical chiral superfields (prepotentials),
\begin{align}\label{a1}
A^{++}(x,\q^+,u)\,, \qquad A^+_\da(x,\q^+,u)\,.
\end{align}
Here $\q^{+ A}=\q^A_\a u^{+\a}$ is a projection of the chiral Grassmann variable  with a  harmonic variable $u^{+\a}$. This commuting spinor variable together with its conjugate $u^{-\a}$ form a matrix of the chiral half $SU(2)_L$ of the Euclidean Lorentz group $SO(4) \sim SU(2)_L \times SU(2)_R$. The harmonic variables $u^\pm$ parametrise the coset space $S^2 \sim SU(2)_L/U(1)$. The superfields \p{a1} are interpreted as infinite harmonic expansions on the sphere, i.e.  homogeneous series in the harmonic variables $u^\pm$ with fixed $U(1)$ charge. For example, in the expansion of $A^+_\da(x,\q^+,u)  = A_{\a\da}(x) u^{+\a} +  A_{\a\b\gamma\da}(x) u^{+\a} u^{+\b} u^{-\gamma} +\ldots + O(\q) $ we find the ordinary gauge field $A_{\a\da}(x)$ and an infinite set of auxiliary higher-spin fields $A_{\a\b\gamma\da}(x), \ldots\ $. Note the absence of the other projection $\q^{- A}=\q^A_\a u^{-\a}$ in \p{a1}. Such superfields are called chiral-analytic. 

The prepotentials have the meaning of the connections for two of the gauge covariant derivatives in the theory, namely
\begin{align}\label{a2}
\nabla^{++} = \pa^{++} + A^{++}\,, \qquad \nabla^{+}_\da = \pa^{+}_\da + A^{+}_\da\,.
\end{align}
Here $\pa^{+}_\da =  u^{+\a}\pa_{\a\da}$ is a projection of the space-time derivative $\pa_x$ while $\pa^{++}= u^{+\a}\pa/\pa u^{-\a}$ is one of the two covariant derivatives on $S^2$.   These derivatives transform with a gauge parameter of the chiral-analytic type,
\begin{align}\label{a3}
\nabla \ \to \   e^{\Lambda(x,\q^+,u)}\ \nabla\ e^{-\Lambda(x,\q^+,u)}\,.
\end{align}
The remaining gauge connections can be constructed from the prepotentials by solving the various super-curvature constraints. In particular, the projected spinor derivative $\pa^+_A =  u^{+\a}\pa/\pa \q^{\a A}$ commutes with the gauge parameter $\Lambda(x,\q^+,u)$, hence it needs no connection,
$\nabla^+_A=  \pa^+_A $.

The action of the theory consists of two terms,
\begin{align}\label{L}
S_{\cN=4\ SYM} = \int du d^4 x d^4\q^+ \ L_{CS}(x,\q^+,u) + \int d^4x d^8\q\ L_Z(x,\q)\,.
\end{align}
The first term in \p{L} is of the Chern-Simons type,
\begin{align}\label{CS}
L_{\rm CS}(x,\q^+,u) =  \tr\; \left(A^{++}\pa^{+\da}A^+_\da-
{1\over 2} A^{+\da}\pa^{++} A^+_\da + A^{++} A^{+\da} A^+_\da\right) 
\end{align}
and it describes the self-dual sector of the theory  \cite{Sokatchev:1995nj}. The second term in \p{L} involves only the prepotential $A^{++}$ in a non-polynomial way \cite{Zupnik:1987vm,Mason:2005zm},
\begin{align}\label{lint}
L_{\rm Z}    = \tr\sum^\infty_{n=2}{(-1)^n\over n} \int du_1\ldots du_n\; {A^{++}(x, \q^+_1,u_1) \ldots A^{++}(x, \q^+_n,u_n) \over
\vev{u^+_1u^+_2} \ldots \vev{u^+_nu^+_1}}\,,
\end{align} 
where $\q^{+A}_i = \q^{\a A} (u_i)^+_{\a}$ with $i=1,\ldots,n$ and $\vev{u_i^+ u_j^+} =u_i^{+\alpha} \ep_{\a\b}\, u^{+\b}_{j}$.  This Lagrangian is local in $(x,\q)$ space but non-local in the harmonic space (each copy of $A^{++}$ depends on its own harmonic variable). The gauge coupling constant $g$   can be restored by redefining $A \to g A$ and $L \to g^{-2}L$.

In this paper we are dealing with form factors, so we need to define the supersymmetric on-shell states. A detailed discussion can be found in \cite{Chicherin:2016qsf}, here we only recall that the super-wave functions of the prepotentials $A$ in the state with (super)momentum $(k,\eta)$ have the form
\begin{align}
\vev{k,\eta|A^{++}(x,\theta^+,u)|0} = \delta^2(k,u) e^{i k x + \vev{k \theta}\eta}\,\,,\qquad\,\vev{k,\eta|A^{+}_{\da}(x,\theta^+,u)|0} =0 
\end{align}
provided we quantise the theory in the light-cone gauge \p{a8}.
The harmonic delta function $\delta^2(k,u)$ identifies the harmonic variable of the field with the chiral spinor momentum, $u^+_\a = k_\a$. Notice that only the prepotential $A^{++}$ has a non-trivial wave function, while $A^+_\da$ does not appear in external states.

\subsection{Chiral Wilson loop  in LHC superspace}

Now, the question arises how to reformulate the Wilson loop \p{W-super}, \p{127} in terms of the prepotentials? The detailed answer is given   in Appendix~\ref{apA}, here we just summarise it.

The chiral lightlike Wilson loop in LHC superspace takes the following form:
\begin{align}\label{2.17}
{\cal W}_n = \frac{1}{N} \tr\prod_{i=1}^n U(x_i,\q_i;  p_{i},  p_{i-1}) E_{i+1,i}\,.
\end{align}
Here  the {so-called}  bilocal bridge 
\begin{align} \label{U'}
U(x,\q;p_2, p_1) = 1 + \sum_{n = 1}^{\infty} (-1)^n \int d u_1 \ldots d u_n
\frac{\vev{p_2 p_1} A^{++}(1)\ldots A^{++}(n)}{\vev{p_2 u_1^+} \vev{u_1^+ u_2^+} \ldots \vev{u_n^+ p_1}} 
\end{align}
resembles  the interaction Lagrangian \p{lint}. The bridges glue together adjacent 
Wilson line segments in \p{2.17}, 
\begin{align}\label{2.18}
  E_{i+1,i} = P \exp\left\{-\frac{i}{2}\int_0^{1} dt \   \tilde p^\da_i A^+_\da\Big(x_i- t  \tilde p_i p_i , \vev{ p_i  \q_i},  \ket{p_i} \Big)  \right\}\,.
\end{align}
We remark that {in the expression for the Wilson loop \p{2.17}}
the prepotential $A^{++}$ appears only at the cusps of the Wilson loop contour via the bilocal bridge $U$ \p{U'}, while the other prepotential $A^{+}_\da$ contributes only through the edges of the contour. 
 
We would like to emphasise that {the definition of the} Wilson loop \p{2.17} differs from the twistor formulation of Mason and Skinner  \cite{Mason:2010yk} {in that it contains the additional Wilson line segments $E_{i+1,i}$ (see the discussion in App.~\ref{aA1}). We believe that the definition
of the Wilson loop in  \cite{Mason:2010yk}  is not gauge invariant and hence it is incomplete. Still, the result  of their calculation of the NMHV Wilson loop is correct, for a reason which will become clear in Section~\ref{sec:nmhv}. However, as we show in this paper, the Wilson line segments in \p{2.17} are indispensable for obtaining a gauge-invariant result for the Wilson loop form factor.} 

 They have the analog of the bilocal bridge $U$ (called `parallel propagator') but not the Wilson line segments $E_{i+1,i}$ (see the discussion in App.~\ref{aA1}). We believe that their definition is incomplete (in fact, not even gauge invariant!). Still, the result  of their calculation of the NMHV Wilson loop is correct, for a reason which will become clear in Section~\ref{sec:nmhv}. However, as we show in this paper, the Wilson line segments in \p{2.17} are indispensable for obtaining a gauge-invariant result for the Wilson loop form factor.

\section{Diagrammatic approach to the duality}

In this section we illustrate the duality \p{duality-gen} in the simplest MHV$\times$MHV case. It corresponds to the first term on the right-hand side of (\ref{W-hat}) which has the lowest Grassmann degree $(\kappa=0,\sigma=0)$. We apply the Feynman rules from Appendix \ref{apB} to the calculation of the Wilson loop form factor  defined in \p{2.17}, in the planar limit and to the lowest order in the coupling  and rederive the result \p{MHV-sum}. This example illustrates
both the graph duality and the simplicity of the LHC computation by
applying the effective rules of Appendix~\ref{sec:effective-rules}.

We end the section by a discussion of the general structure of the non-MHV diagrams.

\subsection{MHV example}
\label{sec:mhv-example}

\begin{figure}[ht!]
\begin{tikzpicture}
  \pgfmathsetmacro{\R}{2} 
  \foreach \x [count=\xi] in {60,30,...,-301}
  {\coordinate[dot] (x\xi) at (\x:\R) ;}
  \foreach \x [count=\xi] in {60,30,...,-301}
  {\coordinate (k\xi) at (\x:.3*\R) ;}
  \draw[double] (x1.center) \foreach \xx in {2,...,12} 
 { -- node[auto,inner sep=1pt]{$p_{\xx}$} (x\xx.center) } -- node[auto]{$p_1$} (x1.center) ; 
  \draw[photon] (x3) -- node[above]{$k_8$}(k1);
  \draw[photon] (x3) -- node[auto]{$k_6$}(k5);
  \draw[photon] (x3) -- (k3);
  \draw[photon] (x6) -- node[right]{$k_5$}(k6);
  \draw[photon] (x7) -- node[auto]{$k_4$}(k7);
  \draw[photon] (x10) -- node[below]{$k_3$}(k8);
  \draw[photon] (x10) -- node[auto]{$k_{2}$}(k11);
  \draw[photon] (x12) -- node[auto]{$k_{1}$}(k12);
  \draw[fill=white] (0,0) circle (.3*\R);
  \draw    (0,0) node  {$\infty$};     
\begin{scope}[xshift=2.5cm] 
 \draw (0,0) node {$\rightarrow$};
 \end{scope}
  \begin{scope}[xshift=5.5cm]
% Define Coords for WL
  \foreach \x [count=\xi] in {60,30,...,-301}
  {\coordinate[dot] (x\xi) at (\x:\R) ;}
  \foreach \x [count=\xi] in {60,30,...,-301}
  {\coordinate (k\xi) at (\x:.3*\R) ;}
  \draw[double] (x1.center) \foreach \xx in {2,...,12} 
  { -- (x\xx.center) } -- (x1.center) ; 
  \draw[photon,thick] (x3) -- (k1);
  \draw[photon,thick] (x3) -- (k5);
  \draw[photon,thick] (x3) -- (k3);
  \draw[photon,thick] (x6) -- (k6);
  \draw[photon,thick] (x7) -- (k7);
  \draw[photon,thick] (x10) -- (k8);
  \draw[photon,thick] (x10) -- (k11);
  \draw[photon,thick] (x12) -- (k12);
  \draw[fill=white] (0,0) circle (.3*\R);
  \draw    (0,0) node  {$\infty$};     
% Coordinates for dual WL
  \foreach \x [count=\xi] in {405,370,350,315,255,195,150,120}
  {\coordinate[dotg] (y\xi) at (\x:.5*\R) ;}
  \foreach \x [count=\xi] in {75,45,...,-316}
  {\coordinate (p\xi) at (\x:1.3*\R) ;}
%draw dual WL
  \draw[double,black!25] (y1.center) \foreach \xx in {2,...,8} 
  { -- (y\xx.center) } -- (y1.center) ; 
\draw[photon,black!25] (y1) -- (p1);
\draw[photon,black!25] (y1) -- (p2);
\draw[photon,black!25] (y1) -- (p3);
\draw[photon,black!25] (y4) -- (p4);
\draw[photon,black!25] (y4) -- (p5);
\draw[photon,black!25] (y4) -- (p6);
\draw[photon,black!25] (y5) -- (p7);
\draw[photon,black!25] (y6) -- (p8);
\draw[photon,black!25] (y6) -- (p9);
\draw[photon,black!25] (y6) -- (p10);
\draw[photon,black!25] (y8) -- (p11);
\draw[photon,black!25] (y8) -- (p12);
  \end{scope}
\begin{scope}[xshift=8cm] 
 \draw (0,0) node {$\rightarrow$};
 \end{scope}
  \begin{scope}[xshift=11cm]
% Coordinates for dual WL
  \foreach \x [count=\xi] in {405,370,350,315,255,195,150,120}
  {\coordinate[dot] (y\xi) at (\x:.5*\R) ;}
  \foreach \x [count=\xi] in {75,45,...,-316}
  {\coordinate (p\xi) at (\x:1.3*\R) ;}
  % draw dual WL
  \draw[double] (y1.center) \foreach \xx [count=\xi] in {8,7,...,1} 
  { --node[auto,inner sep=0pt]{$k_{\xi}$} (y\xx.center) } --
  (y1.center) ; 
\draw[photon] (y1) -- node[auto,inner sep=1pt]{$p_{1}$} (p1);
\draw[photon] (y1) --  node[auto,inner sep=1pt]{$p_{2}$}(p2);
\draw[photon] (y1) --  node[auto,inner sep=1pt]{$p_{3}$}(p3);
\draw[photon] (y4) --  node[auto,inner sep=1pt]{$p_{4}$}(p4);
\draw[photon] (y4) --  node[auto,inner sep=1pt]{$p_{5}$}(p5);
\draw[photon] (y4) --  node[auto,inner sep=1pt]{$p_{6}$}(p6);
\draw[photon] (y5) --  node[auto,inner sep=1pt]{$p_{7}$}(p7);
\draw[photon] (y6) --  node[auto,inner sep=1pt]{$p_{8}$}(p8);
\draw[photon] (y6) --  node[auto,inner sep=1pt]{$p_{9}$}(p9);
\draw[photon] (y6) --  node[auto,inner sep=1pt]{$p_{10}$}(p10);
\draw[photon] (y8) --  node[auto,inner sep=1pt]{$p_{11}$}(p11);
\draw[photon] (y8) --  node[auto,inner sep=1pt]{$p_{12}$}(p12);
  \end{scope}
\end{tikzpicture}

\caption{ 
{The left figure represents a planar Born-level diagram for the Wilson loop form factor $W^{(0,0)}_{12,8}$.
The external particles are coming from infinity which is chosen inside the Wilson loop contour.
The right figure represents a diagram for $W^{(0,0)}_{8,12}$ where the variables specifying the Wilson loop contour
and the external particles are swapped. Here infinity is chosen to lie outside the Wilson loop contour.
In the middle figure the two diagrams are superimposed so that the planar graph duality is manifest.}}\label{fig:3}
\end{figure}
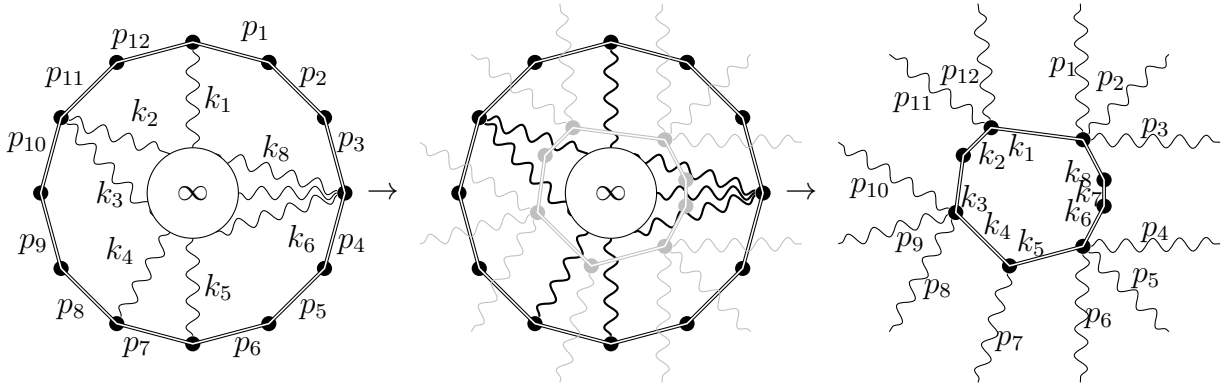

As follows from the definition of the Wilson loop \p{2.17} -- \p{2.18}, to the lowest degree in the Grassmann variables, the Born-level contribution
only comes from diagrams without internal propagators and interaction vertices and with the prepotential  $A^{++}$ replaced by the wave function \p{B10}. Indeed, the propagators of the prepotentials $A^{++}$ and $A^+_{\dot\alpha}$ given by \p{B9}, \p{B12}, \p{B13} and \p{B14} are nilpotent 
(either $\sim\q^4$ or $\sim\eta^4$) and increase the Grassmann degree. This leaves us with only one type of diagram illustrated in Figure~\ref{fig:3}. 

\noindent
Here in the diagram on the left-hand side we draw all external legs inside
the Wilson loop contour. These legs are ordered according to  (\ref{W-def})
and they end at a point that we call `infinity'. This graph contains $n=12$ edges
and $m=8$ external particles and contributes to $F_{12,8}$. In
the second diagram in Figure~\ref{fig:3} we show the planar dual graph
faintly superimposed. For every face of the original graph we draw a
vertex, then we join the vertices up by appropriate edges going
through the boundaries of two faces as described above. This results in 
the third diagram which we recognise as a valid MHV$\times$MHV diagram
contributing to $F_{8,12}$ with all external legs outside the Wilson loop.
 
Let us compute the graph expressions using the simple rules
from Appendix~\ref{sec:effective-rules}. 
First consider the left diagram in Figure~\ref{fig:3}. There are five non-trivial
cusps of the Wilson loop emitting particles. Using~\eqref{B7} and \eqref{B11} we obtain the following contribution to $F_{12,8}$ 
\begin{align}\notag
 F_{12,8} {}&= \frac{e^{ik_1
  x_1+ Q_1 \theta_1}\vev{p_{12}p_{1}}}{\vev{p_1k_1}\vev{k_1p_{12}}}\times
  \frac{e^{i(k_2+k_3)
  x_{11}+ (Q_2+Q_3)\theta_{11}}\vev{p_{10}p_{11}}}{\vev{p_{11}k_2}\vev{k_2k_3}\vev{k_3p_{10}}}
  \\
 {}& \times \frac{e^{i k_4 
  x_8 + Q_4 \theta_8}\vev{p_{7}p_{8}}}{\vev{p_{8}k_4}\vev{k_4p_{7}}}
  \times \frac{e^{ik_5 x_7+Q_5 \theta_7}\vev{p_{6}p_{7}}}{\vev{p_{7}k_5}\vev{k_5p_{6}}}
\times
  \frac{e^{i(k_6+k_7+k_8)
  x_{4}+ (Q_6+Q_7+Q_8)\theta_4}\vev{p_{3}p_{4}}}{\vev{p_{4}k_6}\vev{k_6k_7}\vev{k_7k_8}\vev{k_8p_{3}}}\ .
\end{align}
where $Q_i \theta_j \equiv \eta_{i A} \vev{k_i\theta_j^A}$. The
dependence on the Grassmann variables follows the similar  bosonic variables
exponents.  Substituting $F_{12,8}$ into \eqref{Anm} and (\ref{sum-gen}) we obtain the corresponding 
contribution to $\widehat W_{12,8}$  
\begin{align}\label{eq:8} 
\widehat W_{12,8}= \frac{\vev{k_8
  k_1}\vev{k_1k_2}\vev{k_3k_4}\vev{k_4k_5}\vev{k_5k_6}\times\vev{p_{12}p_1}\vev{p_{10}p_{11}}\vev{p_7p_8}\vev{p_{6}p_{7}}\vev{p_{3}p_4}} {\vev{p_1 k_1}\vev{k_1 p_{12}}\vev{p_{11} k_2}\vev{k_3 p_{10}}\vev{p_8
  k_4}\vev{k_4p_7}\vev{p_{7} k_5}\vev{k_5p_{6}}\vev{p_{4}
    k_6}\vev{k_8p_3 }  }\ .
\end{align}

Let us now look at the right diagram in Figure~\ref{fig:3}. It depends on the variables $(y_j,\psi_j)$ defining the Wilson loop contour and
the variables $(p_i,\omega_i)$ specifying the external particles. 
Using the effective Feynman rules, we obtain the following contribution to $F_{8,12}$: 
\begin{align}
F_{8,12} &=  \frac{e^{i(p_1+p_2+p_3)
  y_1 + (\tilde Q_1+\tilde Q_2+\tilde Q_3)\psi_1}\vev{k_{8}k_{1}}}{\vev{k_8p_3}\vev{p_3p_2}\vev{p_2p_1}\vev{p_1k_{1}}}\times
  \frac{e^{i(p_{11}+p_{12})
  y_{2}+ (\tilde Q_{11}+\tilde Q_{12})\psi_2}\vev{k_{1}k_{2}}}{\vev{k_{1}p_{12}}\vev{p_{12}p_{11}}\vev{p_{11}k_{2}}}\notag
  \\ \notag
&  
  \times \frac{e^{i(p_{8}+p_9+p_{10})
  y_4 + (\tilde Q_8 + \tilde Q_9 +\tilde Q_{10})\psi_4}\vev{k_{3}k_{4}}}{\vev{k_{3}p_{10}}\vev{p_{10}p_{9}}\vev{p_9p_8}\vev{p_8k_4}}
  \times \frac{e^{i p_7 y_5 + \tilde Q_7 \psi_5 }\vev{k_{4}k_{5}}}{\vev{k_{4}p_7}\vev{p_7k_{5}}}
  \\
 {}& 
\times
  \frac{e^{i(p_4+p_5+p_6)
  y_{6}+ (\tilde Q_4 + \tilde Q_5 +\tilde Q_6)\psi_6}\vev{k_{5}k_{6}}}{\vev{k_{5}p_6}\vev{p_6p_5}\vev{p_5p_4}\vev{p_4k_{6}}}\,,
\end{align}
where $\tilde Q_i\psi_j \equiv \omega_{i}^A \vev{p_i \psi_{j A}}$. Substituting this expression into \eqref{Anm} and (\ref{sum-gen})
we find that its contribution to $\widehat W_{8,12}$ is precisely equal to (\ref{eq:8}),
\begin{align}
\widehat W_{8,12}=\widehat W_{12,8}\,.
\end{align} 
This example illustrates the general diagrammatic proof of the duality
in the MHV
case: there are mixed $\langle k_i p_j \rangle$ brackets, common to both the graph and
its dual. Then the missing $\vev{k_i k_j}$ brackets in the denominator on
one side become explicit numerator terms from the Wilson loop vertices
on the other, and vice versa for the $\vev{p_ip_j}$ brackets.    

The exponential factors can be seen to agree in general, also diagrammatically. {Using $k_j=y_j-y_{j+1}$ we find that }
there is an exponent $e^{ i x_i y_j}$ in the left
diagram   if 
and only if the  face $y_j$ has a corner $x_i$. In the dual picture
faces and vertices are swapped, but the result is unchanged. The Grassmann exponents follow the same pattern.

\subsection{Classification of diagrams}

Going beyond the MHV level, we have to consider diagrams containing propagators and interaction vertices. An example
of such a diagram is shown in Figure~\ref{fig:NMHV}. The left diagram contains two propagators connecting cusp points and one propagator connecting a cusp point with a vertex of emission of two particles. The former 
produce a  factor $O(\theta^8)$ and the latter yields a  factor $O(\eta^4)$. As a result, this diagram describes an 
 N${}^2$MHV$\times$NMHV contribution. Similarly, the contribution of the right diagram in Figure~\ref{fig:NMHV} is NMHV$\times$N${}^2$MHV like. In general, a diagram with $\kappa + \sigma$ propagators (dash lines) and $\sigma$ emission
vertices (white dots) gives rise to an N${}^\kappa$MHV$\times$N${}^\sigma$MHV  contribution in the Born approximation.  

The duality essentially works   diagram by   diagram,
although there are a few subtleties which will be discussed in the next
section. Indeed the duality can be seen very straightforwardly at
the diagrammatic level and is essentially a planar graph duality. 

Planarity in this context is slightly non-trivial since on the one
hand we have
external legs coming from infinity and on the other,  position space
propagators between the cusp points. In the planar
limit the only diagrams that survive are those which can be drawn 
with  all external legs and all
internal propagators ``outside'' the Wilson loop contour, without any of the
lines crossing, and with the external legs going to
infinity (see left diagram in Figure~\ref{fig:NMHV}). 
There is an alternative description where we move the source of the
external legs ``infinity'' to a point inside the Wilson loop and
insist that all legs and internal propagators lie {\em inside} the
Wilson 
loop without any lines crossing (see the right diagram in Figure~\ref{fig:NMHV}). 
 
To obtain the  dual of a contributing graph, draw a vertex inside each
face, and 
connect vertices by lines going through the boundaries of the faces:
if the boundary of two faces is a leg (wavy line), connect the corresponding
vertices with a Wilson line (double line), if the boundary is a Wilson line connect
with a leg and if the boundary is a propagator (dash line), connect them
with propagators (see the middle diagram in Figures~\ref{fig:NMHV} and \ref{fig:3}). 

The resulting dual graph will be a valid graph contributing to the
dual Wilson loop form factor (but in the opposite description i.e. if
the original graph had all propagators outside the Wilson loop contour, the
dual graph has all propagators inside). Furthermore, the
expressions for the two graphs are identical after swapping the
variables appropriately.

{Let us note that the case $W_{n,0}^{(\kappa,0)} = W_{0,n}^{(0,\kappa)}$ corresponds to the 
duality between the vacuum expectation value of the $n$-gon Wilson loop and the N$^{\kappa}$MHV $n$-particle amplitude.
The diagrammatic interpretation of this duality in terms of momentum twistor variables has been discussed in \cite{Mason:2010yk}.}

\section{Duality between NMHV$\times$MHV and MHV$\times$NMHV}
\label{sec:nmhv}

We now move on to consider diagrams from the second term in Eq. \p{W-hat}, i.e. $\kappa + \sigma = 1$, which come in two types
$(\kappa,\sigma) = (0,1)$ or $(1,0)$. Each type consists of diagrams with a single propagator. The diagrams of the first type $(0,1)$ have
Grassmann structure $O(\eta^4)$ and contain an emission vertex (white blob). According to the Feynman rules (see  (\ref{B12}) and (\ref{B13})) this 
vertex is connected by a propagator (dash line) either to a cusp point (black blob) or to an edge of the Wilson loop contour (double line). The diagrams of the second type $(1,0)$ have Grassmann structure $O(\theta^4)$. They contain one propagator which is stretched either between two cusps or between a cusp and an edge
 of the Wilson loop contour. We call diagrams where all propagators
end on cusps/vertices   ``cusp diagrams''. Diagrams involving
propagators ending  on edges are called ``edge diagrams''.%
\footnote{Note that only one end of a propagator can be on a Wilson
  loop edge, since there is no propagator between two prepotentials  $A_{\dot
    \alpha}^+$, see Appendix \p{apB}.}

In this section we carefully examine the NMHV-like case
by focussing on the sector of a general  Feynman diagram involving
the propagator. We first examine the cusp diagrams before turning  
to the edge diagrams, show how spurious poles cancel and in the
process allow the duality to hold in a surprising and non-trivial fashion.

\subsection{Cusp diagrams}

First we consider the diagrams contributing to the form factor $W_{n,m}^{(1,0)}$, Eq. \p{W-hat} in the Born approximation. These diagrams contain $n$ cusps (black blobs),
$m$ external states (wave lines) and one propagator (dash line).
Focussing on the part of the diagram  containing
the 
propagator we have \footnote{We use this diagram to
illustrate the general case. The most  generic  diagram would have
an arbitrary number of legs (or none) in place of $k_1$ and $k_2$. }
\begin{align}
\begin{tikzpicture}
  \coordinate[dot,label={$x_2$}](x2);
  \coordinate[dot,below=2.7cm of x2,label=below:{$x_i$}](xi);
  \coordinate[dot,below left=1cm of x2](x1);
  \coordinate[below=.2cm of x1](x1ext){};
  \coordinate[dot,below right=1cm of x2](x3);
  \coordinate[below=.2cm of x3](x3ext){};
  \coordinate[dot,above left=1cm of xi](xip1);
  \coordinate[above=.2cm of xip1](xip1ext){};
  \coordinate[dot,above right=1cm of xi](xim1);
  \coordinate[above=.2cm of xim1](xim1ext){};
  \node[right=.3cm of x1,inner sep=-2pt](k1){$k_1$};
  \node[right=.3cm of xip1,inner sep=-2pt](k2){$k_2$};
  \node[below=0cm of x1]{\vdots};
  \node[below=0cm of x3]{\vdots};
  \draw[very thick,dashed](x2) --node[right]{} (xi);
  \draw[double] (x1)--node[auto]{$p_1$} (x2)--node[auto]{$p_2$} (x3);
  \draw[double] (xim1)--node[auto]{$p_{i-1}$} (xi)--node[auto]{$p_i$}
  (xip1);
  \draw[double] (x1)--(x1ext);
  \draw[double] (x3)--(x3ext);
  \draw[double] (xim1)--(xim1ext);
  \draw[double] (xip1)--(xip1ext);
  \draw[photon] (k1) -- (x2);
  \draw[photon] (k2) -- (xi);
\end{tikzpicture}
\raisebox{1.7cm}
{$\displaystyle = \int \frac{d^4 q}{4\pi^2}\, e^{i q x_{2i}}
  \frac{e^{i(k_1 x_2+k_2 x_i)}\vev{p_1p_2}\vev{p_{i-1}p_i}\delta^4\big(\bra{\theta_{2i}}q|\xi]\big)}{q^2\vev{p_1k_1}
 \bra{k_1}q|\xi][\xi|q\ket{p_2} 
 \bra{p_{i-1}}q|\xi][\xi|q\ket{k_2}\vev{k_2p_i} }\times \dots$
  }\label{eq:4}
  \end{align}
where $q$ is the momentum that flows through the dash  line.
Here and in all expressions below we drop the exponential
dependence on the Grassmann variables $e^{\eta_1\vev{k_1\theta_2}+ \eta_2\vev{k_2\theta_i}}$ which follows the similar
exponential factor of the bosonic variables. The dots on the right-hand side of \p{eq:4} denote the contribution of the remaining
cusps which are the same as in the MHV case.

Now consider cusp diagrams contributing to 
$W_{m,n}^{(0,1)}$. Again focussing on the piece containing the
propagator, the dual to the above diagram is:
\begin{align}
\begin{tikzpicture}
  \coordinate[dot,label=below:{$y_2$}](y2);
  \coordinate[dotw,right=2.5cm of y2](y);
  \node[above right=1.5cm of y2]{$x_2$};
  \node[below right=1.5cm of y2]{$x_i$};
  \coordinate[dot,below left=1cm of y2](y1);
  \coordinate[dot,above left=1cm of y2](y3);
  \node[above right=1cm of y](p2){$p_2$};
  \node[below right=1cm of y](pim1){$p_{i-1}$};
  \node[above right=.7cm of y3](p1){$p_1$};
  \node[below right=.7cm of y1](pi){$p_i$};
  \node[below=0.2cm of p2]{\vdots};
  \coordinate[above left=.2cm of y1](y1ext);
  \coordinate[below left=.2cm of y3](y3ext);
 \node[below =0.1cm of y3ext]{\vdots};
  \draw[very thick,dashed](y2) -- (y);
  \draw[double] (y1ext) -- (y1)--node[auto,inner sep=0cm]{$k_2$}
  (y2)--node[auto,inner sep=0cm]{$k_1$} (y3)--(y3ext); 
  \draw[photon] (y) -- (p2);
  \draw[photon] (y) -- (pim1);
  \draw[photon] (y3) -- (p1);
  \draw[photon] (y1) -- (pi);
\end{tikzpicture}
\raisebox{1.7cm}{$ \displaystyle =
  \frac{e^{-i x_{2i}y_2}\vev{k_1k_2}\delta^4\big(\bra{\theta_{2i}}x_{2i}|\xi]\big)}{x_{2i}^2\vev{p_1k_1}
 \bra{k_1}x_{2i}|\xi][\xi|x_{2i}\ket{p_2} 
 \bra{p_{i-1}}x_{2i}|\xi][\xi|x_{2i}\ket{k_2}\vev{k_2p_i} }\times
  \dots$  }\label{eq:3}
  \end{align}
Here we have replaced the momentum through the propagator $p_2+\dots
+p_{i-1}$ by the dual variable $x_{2i}$  and similarly the
supermomentum by $\theta_{2i}$.  

The expressions \eqref{eq:4} and \eqref{eq:3} depend on the gauge fixing spinor $\xi$ (see Eq.~\p{a8}) which generate spurious complex
poles, e.g. $ \bra{k_1}q|\xi]=0$ in  \eqref{eq:3}. Below we demonstrate that the $\xi$-dependence as well as the spurious poles cancel in the sum of all diagrams. 

Notice that the two expressions \eqref{eq:4} and \eqref{eq:3} look
very similar. In fact, if we replaced $q$ in~\eqref{eq:4} by $x_{2i}$,
then the integrand of the Fourier integral would be identical to the
expression in~\eqref{eq:3} (up to $\vev{k_r k_{r+1}}$ and $\vev{p_r p_{r+1}}$ factors
which we expect from the duality relations (\ref{Anm}) and (\ref{duality})).\footnote{We have displayed only part of the exponential factors in \p{eq:4} and \p{eq:3}, the rest goes into the ellipsis. To compare them, in \p{eq:4} we rewrite $k_1 x_2 + k_2 x_i = -y_2 x_{2i} + \ldots$, as in \p{eq:3}. \label{foot2}} Indeed if we were allowed to
perform the Fourier transform in \eqref{eq:4} in  Euclidean
space, then we would simply replace $q$ by $x_{2i}$ everywhere according to Eq. \p{412}. 
Thus {\em{if}} we were in Euclidean space,
\eqref{eq:4} and \eqref{eq:3} would give   identical expressions (up
to the Parke-Taylor factors) leading to the required
duality between diagrams. At the moment however we cannot yet justify Wick
rotation to Euclidean space as \eqref{eq:4} has spurious complex poles in
$q$-space preventing this. It will become possible after we take into account the edge diagrams.

\subsection{Edge diagrams}
\label{sec:edge-diaygrams}

Besides the cusp diagrams we also have edge diagrams with propagators
ending on (and being integrated along) edges of the Wilson loop contour. These
appear in both $W_{n,m}^{(1,0)}$ and $W_{m,n}^{(0,1)}$.

%\subsection*{NMHV$\times$MHV sector}

An example of  such a contribution to $W_{n,m}^{(1,0)}$ is
\begin{align}
\begin{tikzpicture}[>=stealth']
  \coordinate[dot,label={$x_2$}](x2);
  \coordinate[below right=.5cm of x2](t);
  \coordinate[dot,below=2.7cm of x2,label=below:{$x_i$}](xi);
  \coordinate[dot,below left=1cm of x2](x1);
  \coordinate[below=.2cm of x1](x1ext){};
  \coordinate[dot,below right=1cm of x2](x3);
  \coordinate[below=.2cm of x3](x3ext){};
  \coordinate[dot,above left=1cm of xi](xip1);
  \coordinate[above=.2cm of xip1](xip1ext){};
  \coordinate[dot,above right=1cm of xi](xim1);
  \coordinate[above=.2cm of xim1](xim1ext){};
  \node[right=.3cm of x1,inner sep=-2pt](k1){$k_1$};
  \node[right=.3cm of xip1,inner sep=-2pt](k2){$k_2$};
  \node[below=0cm of x1]{\vdots};
  \node[below=0cm of x3]{\vdots};
  \draw[<-, very thick,dashed](t) --node[right]{} (xi);
  \draw[double] (x1)--node[auto]{$p_1$} (x2)--node[auto]{$p_2$} (x3);
  \draw[double] (xim1)--node[auto]{$p_{i-1}$} (xi)--node[auto]{$p_i$}
  (xip1);
  \draw[double] (x1)--(x1ext);
  \draw[double] (x3)--(x3ext);
  \draw[double] (xim1)--(xim1ext);
  \draw[double] (xip1)--(xip1ext);
  \draw[photon] (k1) -- (x2);
  \draw[photon] (k2) -- (xi);
\end{tikzpicture}
\raisebox{1.7cm}{$ \displaystyle = \int \frac{d^4 q}{4\pi^2} 
  \frac{\big(e^{iq x_{2i}}-e^{iq x_{3i}}\big)e^{i(k_1 x_2+k_2 x_i)}\vev{p_1p_2}\vev{p_{i-1}p_i}[p_2 \xi]\delta^4\big(\vev{\theta_{2i}p_2}\big)}{\vev{p_1k_1}\vev{k_1p_2}
 \bra{p_2}q|\xi] [p_2|q\ket{p_2}  
 \vev{p_{i-1}p_2}\vev{p_2k_2}\vev{k_2p_i} }\times \dots$
  }\label{eq:4b}
  \end{align}
We note here that, again if we could perform the Fourier integration
in Euclidean space, we would get two terms coming from the two exponents in the first factor.
The first term is obtained by replacing $q$ in the integrand by $x_{2i}$ 
and the second one replacing $q$ by $x_{3i}$, see Eq. \p{412}. But these two
terms are equal (and opposite) since  $x_{3i}= x_{2i} + x_{32}= x_{2i}
-p_2$ and $q$ appears everywhere contracted with a $p_2$. Thus as a
Euclidean Fourier integral we get from \p{eq:4b}  a vanishing result and indeed we will
find no corresponding dual diagram as we will discuss shortly.
 
However,  as a Fourier integral in Minkowski space, \p{eq:4b}  is non-vanishing
and plays  an important role in the cancellation of  spurious poles which
ultimately allows for the Wick rotation to Euclidean space. We will take
a closer look at spurious pole cancellation in the next subsection,
but for the
moment we leave all the diagrams contributing to $W_{n,m}^{(1,0)}$
in the form of Fourier integrals.

%\subsection*{MHV$\times$NMHV sector}

There are also edge diagrams contributing to $W_{m,n}^{(0,1)}$ for
example
  \begin{align}
\begin{tikzpicture}[>=stealth']
  \coordinate[dot,label=below:{$y_2$}](y2);
  \node[above right=1.5cm of y2]{$x_2$};
  \node[below right=1.5cm of y2]{$x_i$};
  \coordinate[dot,below left=1cm of y2](y1);
  \coordinate[dot,above left=1cm of y2](y3);
  \coordinate[above left=.5cm of y2](t);
  \coordinate[dotw,right=2.5cm of y2](y);
  \node[above right=1cm of y](p2){$p_2$};
  \node[below right=1cm of y](pim1){$p_{i-1}$};
  \node[above right=.7cm of y3](p1){$p_1$};
  \node[below right=.7cm of y1](pi){$p_i$};
  \node[below=0.2cm of p2]{\vdots};
  \coordinate[above left=.2cm of y1](y1ext);
  \coordinate[below left=.2cm of y3](y3ext);
 \node[below =0.1cm of y3ext]{\vdots};
  \draw[<-,very thick,dashed](t) -- (y);
  \draw[double] (y1ext) -- (y1)--node[auto,inner sep=0cm]{$k_2$}
  (y2)--node[auto,inner sep=0cm]{$k_1$} (y3)--(y3ext); 
  \draw[photon] (y) -- (p2);
  \draw[photon] (y) -- (pim1);
  \draw[photon] (y3) -- (p1);
  \draw[photon] (y1) -- (pi);
\end{tikzpicture}
\raisebox{1.7cm}{$ \displaystyle =
  \frac{(e^{-i x_{2i}y_2}-e^{-i
    x_{2i}y_1})[\xi k_1]  \vev{k_1k_2}\delta^4\big(\vev{\theta_{2i} k_1}\big)}{\vev{p_1k_1}
 \bra{k_1}x_{2i}|\xi]\bra{k_1}x_{2i}|k_1]\vev{k_1p_2} 
 \vev{p_{i-1}k_1}\vev{k_2p_i} }\times
  \dots$  }\label{eq:3b}
  \end{align}
This type of diagram also plays a crucial role in the cancellation of
spurious poles. Again there is no corresponding dual
diagram. Intriguingly however, if we used the duality to translate this
diagram into an expression contributing to the dual diagram $W_{n,m}^{(1,0)}$ we
would obtain an integral which evaluates to zero in Minkowski space. However
this vanishing Minkowski  integral 
plays a crucial role in the cancellation of leftover spurious
poles of $W_{n,m}^{(1,0)}$ at the level of the Fourier integrand. Armed with
these additional terms we then  have a Fourier integrand  with no
remaining 
spurious poles, and we can hence Wick rotate and do the Fourier
integration. After this has been performed these ``fake" terms
contribute a non-vanishing result! We will see all this more
explicitly in 
the next two subsections.

{We end this subsection by a comment on the vacuum expectation value of the lightlike Wilson loop $\vev{0|\mathcal W_n|0}$  at NMHV level. It corresponds to diagrams of the type \p{eq:4} and \p{eq:4b} {\it without external legs}. In \cite{Mason:2010yk} the same object has been calculated in the twistor framework using the incomplete expression \p{MS} for the Wilson loop. Evidently,  this can account only for the cusp diagrams but the edge diagrams are missing. Nevertheless, the final result is correct.  {The reason for this is that} the authors of \cite{Mason:2010yk} work in Euclidean space where the missing edge diagrams vanish, as we have shown. However, the edge diagrams become very important when we consider the form factors of the Wilson loop  \p{W-def0}.}

\subsection{Spurious pole cancellation}
\label{sec:spur-pole-canc}

In this subsection we take the  generic diagrams \eqref{eq:4} and
\eqref{eq:3} contributing to $W_{n,m}^{(1,0)}$ and $W_{m,n}^{(0,1)}$, respectively,
examine their spurious poles and show how they cancel. We begin
with   $W_{m,n}^{(0,1)}$   since there things are more straightforward.

\subsection*{MHV$\times$NMHV sector}
\label{}

Diagram~\eqref{eq:3} contains four
spurious poles. We consider each pole in turn.

It is convenient to associate the spurious poles of \eqref{eq:3} with the four angles formed by the propagator
and the two pairs of lines $(k_1,k_2)$ and $(p_2, p_{i-1})$ attached to each end.   First we consider the pole at $[\xi|x_{2i}\ket{p_2}=0$ (associated with
the upper  right angle). By setting
$|\xi]=x_{2i}\ket{p_2}$ it is straightforward to find the residue of \eqref{eq:3} at
this pole. We can then check that the residue   is
cancelled by the pole of a nearby diagram with the leg $p_2$ attached
to the other end of the propagator.
Diagrammatically, displaying the residue by filling in the associated
angle we have

\begin{align}
\label{eq:4d}
\begin{tikzpicture}[x=.7cm,y=.7cm]
%[scale=0.6, every node/.style={scale=0.5}]
  \coordinate[dot,label=below:{}](y2);
  \coordinate[right=2.5 of y2,dotw](y);
  \node[above right=0.3 of y](ang1){};
  \node[left=0.3 of y](ang2){};
  \node[above right=1.5 of y2]{};
  \node[below right=1.5 of y2]{};
  \coordinate[dot,below left=1 of y2](y1);
  \coordinate[dot,above left=1 of y2](y3);
  \node[above right=1 of y](p2){$p_2$};
  \node[below=.3 of p2](p3){$p_3$};
  \node[below right=1 of y](pim1){};
  \node[above right=.7 of y3](p1){};
  \node[below right=.7 of y1](pi){};
  \node[below=-.3 of p3]{\vdots};
  \coordinate[above left=.2 of y1](y1ext);
  \coordinate[below left=.2 of y3](y3ext);
 \node[below =0.1 of y3ext]{\vdots};
 \begin{pgfonlayer}{background}  \fill[red]
   (y.center)--(ang1.center)--(ang2.center);
 \end{pgfonlayer}
  \draw[very thick,dashed](y2) -- (y);
  \draw[double] (y1ext) -- (y1)--node[auto,inner sep=0]{}
  (y2)--node[auto,inner sep=0]{} (y3)--(y3ext); 
  \draw[photon] (y) -- (p2);
  \draw[photon] (y) -- (p3);
  \draw[photon] (y) -- (pim1);
  \draw[photon] (y3) -- (p1);
  \draw[photon] (y1) -- (pi);
  \node[below right=.6 of p2]{$+$};
  \begin{scope}[xshift=5cm]
    \coordinate[dot,label=below:{}](y2);
  \coordinate[right=2.5 of y2,dotw](y);
  \node[above right=1.5 of y2]{};
  \node[below right=1.5 of y2]{};
  \node[above right=0.3 of y2](ang1){};
  \node[right=0.3 of y2](ang2){};
  \coordinate[dot,below left=1 of y2](y1);
  \coordinate[dot,above left=1 of y2](y3);
  \node[above right=1 of y](p3){$p_3$};
  \node[above right=1 of y2](p2){$p_2$};
  \node[below right=1 of y](pim1){};
  \node[above right=.7 of y3](p1){};
  \node[below right=.7 of y1](pi){};
  \node[below=0.2 of p3]{\vdots};
  \coordinate[above left=.2 of y1](y1ext);
  \coordinate[below left=.2 of y3](y3ext);
 \node[below =0.1 of y3ext]{\vdots};
 \begin{pgfonlayer}{background}  \fill[red]
   (y2.center)--(ang1.center)--(ang2.center);
 \end{pgfonlayer}
  \coordinate[right=2.5 of y2,dotw](y);
  \draw[very thick,dashed](y2) -- (y);
  \draw[double] (y1ext) -- (y1)--node[auto,inner sep=0]{}
  (y2)--node[auto,inner sep=0]{} (y3)--(y3ext); 
  \draw[photon] (y) -- (p3);
  \draw[photon] (y2) -- (p2);
  \draw[photon] (y) -- (pim1);
  \draw[photon] (y3) -- (p1);
  \draw[photon] (y1) -- (pi);
  \node[below right=.6 of p3]{$=0$};
  \end{scope}
\end{tikzpicture}
\end{align}
The pole at $\bra{p_{i-1}}x_{2i}|\xi]=0$ is cancelled by a
very similar mechanism, i.e. the nearby diagram with leg $p_{i-1}$
attached to the other end of the propagator cancels this pole. 

Now we consider the pole at $\bra{k_1}x_{2i}|\xi]=0$ (associated with
the upper left angle). By setting
$|\xi]=x_{2i}\ket{k_1}$ we can obtain its residue and it is
straightforward to check that this cancels the corresponding residue
from the first term of the nearby edge diagram~\eqref{eq:3b}. The
second term of the edge diagram then cancels a spurious pole from
another nearby cusp diagram.
Diagrammatically: 
  \begin{align}
\label{eq:5}
\begin{tikzpicture}[>=stealth',x=.7cm,y=.7cm]
  \coordinate[dot,label=below:{}](y2);
  \coordinate[right=2.5 of y2,dotw](y);
  \node[above right=1.5 of y2]{};
  \node[below right=1.5 of y2]{};
  \coordinate[dot,below left=1 of y2](y1);
  \coordinate[dot,above left=1 of y2](y3);
  \coordinate[above left=.4 of y2](ang1);
  \coordinate[right=.4 of y2](ang2);
  \node[above right=1 of y](p2){};
  \node[below right=1 of y](pim1){};
  \node[above right=.7 of y3](p1){};
  \node[below right=.7 of y1](pi){};
  \node[below=0.01 of p2]{\vdots};
  \coordinate[above left=.2 of y1](y1ext);
  \coordinate[below left=.2 of y3](y3ext);
 \node[below =0.01 of y3ext]{\vdots};
  \draw[very thick,dashed](y2) -- (y);
  \draw[double] (y1ext) -- (y1)--node[auto,inner sep=0]{}
  (y2)--node[auto,inner sep=0]{} (y3)--(y3ext); 
  \draw[photon] (y) -- (p2);
 \begin{pgfonlayer}{background}  \fill[red]
   (y2.center)--(ang1.center)--(ang2.center);
 \end{pgfonlayer}
  \draw[photon] (y) -- (pim1);
  \draw[photon] (y3) -- (p1);
  \draw[photon] (y1) -- (pi);
  \node[below right=.3 of p2]{$+$};
\begin{scope}[xshift=4.5cm]
  \coordinate[dot,label=below:{}](y2);
  \node[above right=1.5 of y2]{};
  \node[below right=1.5 of y2]{};
  \coordinate[dot,below left=1 of y2](y1);
  \coordinate[dot,above left=1 of y2](y3);
  \coordinate[above left=.5 of y2](t);
  \coordinate[right=2.5 of y2,dotw](y);
  \node[above right=1 of y](p2){};
  \node[below right=1 of y](pim1){};
  \node[above right=.7 of y3](p1){};
  \node[below right=.7 of y1](pi){};
  \node[below=0.01 of p2]{\vdots};
  \coordinate[above left=.2 of y1](y1ext);
  \coordinate[below left=.2 of y3](y3ext);
 \node[below =0.01 of y3ext]{\vdots};
  \draw[<-,very thick,dashed](t) -- (y);
  \coordinate (ang1) at ($(t)!0.2!(y)$);
  \coordinate[above left =.3 of t](ang2);
  \coordinate[below right =.3 of t](ang3);
 \begin{pgfonlayer}{background}  \fill[red]
   (ang1.center)--(ang2.center)--(ang3.center);
 \end{pgfonlayer}  
  \draw[double] (y1ext) -- (y1)--node[auto,inner sep=0]{}
  (y2)--node[auto,inner sep=0]{} (y3)--(y3ext); 
  \draw[photon] (y) -- (p2);
  \draw[photon] (y) -- (pim1);
  \draw[photon] (y3) -- (p1);
  \draw[photon] (y1) -- (pi);
  \node[below right=.3 of p2]{$+$};
\end{scope}
\begin{scope}[xshift=9cm]
    \coordinate[dot,label=below:{}](y2);
  \coordinate[right=2.5 of y2,dotw](y);
  \node[above right=1.5 of y2]{};
  \node[below right=1.5 of y2]{};
  \coordinate[dot,below left=1 of y2](y1);
  \coordinate[dot,above left=1 of y2](y3);
  \node[above right=1 of y](p2){};
  \node[below right=1 of y](pim1){};
  \node[above right=.7 of y3](p1){};
  \node[below right=.7 of y1](pi){};
  \node[below=0.01 of p2]{\vdots};
  \coordinate[above left=.2 of y1](y1ext);
  \coordinate[below left=.2 of y3](y3ext);
 \node[below =0.01 of y3ext]{\vdots};
  \draw[very thick,dashed](y3) -- (y);
  \coordinate (ang1) at ($(y3)!0.2!(y)$);
  \coordinate[below right =.3 of y3](ang2);
 \begin{pgfonlayer}{background}  \fill[red]
   (y3.center)--(ang1.center)--(ang2.center);
 \end{pgfonlayer}  
 \draw[double] (y1ext) -- (y1)--node[auto,inner
 sep=0]{} 
  (y2)--node[auto,inner sep=0]{} (y3)--(y3ext); 
  \draw[photon] (y) -- (p2);
  \draw[photon] (y) -- (pim1);
  \draw[photon] (y3) -- (p1);
  \draw[photon] (y1) -- (pi);
  \node[below right=.3 of p2]{$=0$};
\end{scope}
\end{tikzpicture}
\end{align}
The spurious pole of~\eqref{eq:3} at $[\xi|x_{2i}\ket{k_2}=0$  is
cancelled by a similar mechanism (essentially the above picture
reflected in the horizontal axis.)

We have thus cancelled all spurious poles of the diagram~\eqref{eq:3}
using nearby diagrams. Of course each new diagram (apart from the edge
diagram) will introduce new spurious poles, which are then cancelled
by further nearby diagrams etc. In this way we see that all spurious
poles are cancelled up to certain special boundary cases which we
consider in Appendix~\ref{sec:boundary-cases}.

\subsection*{NMHV$\times$MHV sector}
\label{}

We now consider the spurious poles of $W_{n,m}^{(1,0)}$ taking the diagram
in~\eqref{eq:4} as a suitably generic  example.  Again there are four spurious poles which we consider in turn.

First consider the pole at $[\xi|q\ket{p_2}=0$. 
This is shared by
the nearby edge diagram 
\eqref{eq:4b}. By considering the limit $[\xi|
\rightarrow \bra{p_2}q$ we can find the residue at this pole of both 
\eqref{eq:4} and
\eqref{eq:4b}. We find that the pole of the first term
of~\eqref{eq:4b} exactly cancels the pole of~\eqref{eq:4}. The second
term of~\eqref{eq:4b} on the other hand cancels a pole of a nearby
cusp diagram. Diagrammatically, denoting the pole of a particular
diagram by filling in the angle between the relevant propagator and
neighbouring edge we find
\begin{align}
\begin{tikzpicture}[>=stealth']
  \coordinate[dot,label={$x_2$}](x2);
  \coordinate[below=.4cm of x2](ang1);
  \coordinate[below right =.4cm of x2](ang2);
 \begin{pgfonlayer}{background}  \fill[red]
   (x2.center)--(ang1.center)--(ang2.center);
 \end{pgfonlayer}
\coordinate[dot,below=2.7cm of x2,label=below:{$x_i$}](xi);
  \coordinate[dot,below left=1cm of x2](x1);
  \coordinate[below=.2cm of x1](x1ext){};
  \coordinate[dot,below right=1cm of x2](x3);
  \coordinate[below=.2cm of x3](x3ext){};
  \coordinate[dot,above left=1cm of xi](xip1);
  \coordinate[above=.2cm of xip1](xip1ext){};
  \coordinate[dot,above right=1cm of xi](xim1);
  \coordinate[above=.2cm of xim1](xim1ext){};
  \node[right=.3cm of x1,inner sep=-2pt](k1){$k_1$};
  \node[right=.3cm of xip1,inner sep=-2pt](k2){$k_2$};
  \node[below=0cm of x1]{\vdots};
  \node[below=0cm of x3]{\vdots};
  \draw[very thick,dashed](x2) --node[right]{} (xi);
  \draw[double] (x1)--node[auto]{$p_1$} (x2)--node[auto]{$p_2$} (x3);
  \draw[double] (xim1)--node[auto]{$p_{i-1}$} (xi)--node[auto]{$p_i$}
  (xip1);
  \draw[double] (x1)--(x1ext);
  \draw[double] (x3)--(x3ext);
  \draw[double] (xim1)--(xim1ext);
  \draw[double] (xip1)--(xip1ext);
  \draw[photon] (k1) -- (x2);
  \draw[photon] (k2) -- (xi);
  \node[below right=.4cm of x3]{$+$};
  \begin{scope}[xshift=3cm]
    \coordinate[dot,label={$x_2$}](x2); 
    \coordinate[below right=.5cm
    of x2](t); 
    \coordinate[dot,below=2.7cm of
    x2,label=below:{$x_i$}](xi); \coordinate[dot,below left=1cm of
    x2](x1); \coordinate[below=.2cm of x1](x1ext){};
    \coordinate[dot,below right=1cm of x2](x3); \coordinate[below=.2cm
    of x3](x3ext){}; \coordinate[dot,above left=1cm of xi](xip1);
    \coordinate[above=.2cm of xip1](xip1ext){}; \coordinate[dot,above
    right=1cm of xi](xim1); \coordinate[above=.2cm of
    xim1](xim1ext){}; \node[right=.3cm of x1,inner
    sep=-2pt](k1){$k_1$}; \node[right=.3cm of xip1,inner
    sep=-2pt](k2){$k_2$}; \node[below=0cm of x1]{\vdots};
    \node[below=0cm of x3]{\vdots}; 
    \draw[<-,very thick,dashed](t)
    --node[right]{} (xi); 
    \coordinate (ang1) at ($(t)!0.2!(xi)$);
  \coordinate[above left =.3cm of t](ang2);
 \coordinate[below right =.3cm of t](ang3);
  \begin{pgfonlayer}{background}
\fill[red] (ang3.center)--(ang1.center)--(ang2.center);
\end{pgfonlayer}
    \draw[double] (x1)--node[auto]{$p_1$}
    (x2)--node[auto]{$p_2$} (x3); \draw[double]
    (xim1)--node[auto]{$p_{i-1}$} (xi)--node[auto]{$p_i$} (xip1);
    \draw[double] (x1)--(x1ext); \draw[double] (x3)--(x3ext);
    \draw[double] (xim1)--(xim1ext); \draw[double] (xip1)--(xip1ext);
    \draw[photon] (k1) -- (x2); \draw[photon] (k2) -- (xi);
  \node[below right=.4cm of x3]{$+$};
  \end{scope}
  \begin{scope}[xshift=6cm]
      \coordinate[dot,label={$x_2$}](x2);
  \coordinate[dot,below=2.7cm of x2,label=below:{$x_i$}](xi);
  \coordinate[dot,below left=1cm of x2](x1);
  \coordinate[below=.2cm of x1](x1ext){};
  \coordinate[dot,below right=1cm of x2](x3);
  \coordinate (ang1) at ($(x3)!0.2!(xi)$);
  \coordinate[above left =.3cm of x3](ang2);
  \begin{pgfonlayer}{background}
\fill[red] (x3.center)--(ang1.center)--(ang2.center);
\end{pgfonlayer}
\coordinate[below=.2cm of x3](x3ext){};
  \coordinate[dot,above left=1cm of xi](xip1);
  \coordinate[above=.2cm of xip1](xip1ext){};
  \coordinate[dot,above right=1cm of xi](xim1);
  \coordinate[above=.2cm of xim1](xim1ext){};
  \node[right=.3cm of x1,inner sep=-2pt](k1){$k_1$};
  \node[right=.3cm of xip1,inner sep=-2pt](k2){$k_2$};
  \node[below=0cm of x1]{\vdots};
  \node[below=0cm of x3]{\vdots};
  \draw[very thick,dashed](x3) --node[right]{} (xi);
  \draw[double] (x1)--node[auto]{$p_1$} (x2)--node[auto]{$p_2$} (x3);
  \draw[double] (xim1)--node[auto]{$p_{i-1}$} (xi)--node[auto]{$p_i$}
  (xip1);
  \draw[double] (x1)--(x1ext);
  \draw[double] (x3)--(x3ext);
  \draw[double] (xim1)--(xim1ext);
  \draw[double] (xip1)--(xip1ext);
  \draw[photon] (k1) -- (x2);
  \draw[photon] (k2) -- (xi);
  \node[below right=.4cm of x3]{$=0$};
  \end{scope}
\end{tikzpicture}\label{eq:4c}
\end{align}
The  spurious pole of~\eqref{eq:4} at $[\xi|q\ket{p_{i-1}}=0$ is
cancelled similarly (by essentially a reflection of~\eqref{eq:4c} about
the horizontal axis).
 
However now consider the spurious pole at $\bra{k_1}q|\xi]=0$. There is
no nearby diagram which can cancel this spurious pole at the level of
the integrand. One is tempted
to consider the diagram with  leg $k_1$ shifted down so that it is
attached to vertex $x_i$, but the two diagrams have different
exponential factors, $e^{ik_1x_2}$ and $e^{ik_1x_i}$, preventing the 
cancellation. At first sight, this leads to an apparent breakdown of gauge invariance in
this sector!

The solution to this puzzle in fact comes from examining the
corresponding cancellation occurring in the dual picture described above. 
There we had a corresponding pole at $\bra{k_1}x_{2i}|\xi]=0$ which
was cancelled by the edge diagram~\eqref{eq:3b}. Translating this via
the duality%
\footnote{More precisely, to obtain this dual expression as  a Fourier
integral, we take the expression~\eqref{eq:3b}, replace $x_{2i}$ with
$q$ in the rational terms, leaving the exponential terms as they are,
multiply by $e^{i q x_{2i}}$ and integrate over $d^4q$. The careful
reader will note that the exponents of \eqref{eq:3b} and~\eqref{eq:6}
are different. In fact the two displayed exponentials are related via  $(e^{-i x_{2i}y_2}-e^{-i
    x_{2i}y_1})=e^{-iy_1 x_1} (e^{i(k_1 x_2+k_2
    x_i)}-e^{i(k_1+k_2)x_i})e^{iy_3x_i}$ and we have simply absorbed
  the factors $e^{-iy_1 x_1}e^{iy_3x_i}$ into the ellipsis. If the
  full exponential 
factors for the entire expressions were written out in both cases they
  would precisely agree (see footnote \ref{foot2}).  }
 suggests we consider the expression 
\begin{align}
  \label{eq:6}
I:=\int \frac{d^4q}{4\pi^2}  e^{iq x_{2i}} \frac{(e^{i(k_1x_2+k_2 x_i)}-e^{i(k_1+k_2)x_i})[\xi k_1]  \vev{p_1p_2}\delta^4\big(\vev{\theta_{2i} k_1}\big)}{\vev{p_1k_1}
 \bra{k_1}q|\xi]\bra{k_1}q|k_1]\vev{k_1k_2}\vev{k_1p_2} 
 \vev{p_{i-1}k_1}\vev{k_2p_i} }\times
  \dots  = 0  \ .
\end{align}
Firstly we note that as an integral in Minkowski space $I$ vanishes and thus we are perfectly at liberty to add 
the integral \p{eq:6} to $W_{n,m}^{(1,0)}$.
The identity $I=0$ can be seen straightforwardly by
observing that the contributions of the two terms inside the parentheses cancel against each other 
after shifting $q \rightarrow q+k_1$ in the second term (since $\bra{k_1}q = \bra{k_1}
(q+k_1)$).

Secondly we note that the integrand of \p{eq:6} is precisely what is
needed to 
recover gauge invariance of the integrand of $W_{n,m}^{(1,0)}$: the spurious pole of the
first term at 
$\bra{k_1}q|\xi]=0$ cancels the spurious pole of the diagram~\eqref{eq:4}.
Furthermore the other term similarly cancels the pole of the diagram
with $k_1,k_2$ both coming from point $x_i$. Thus, for the relevant poles  we
have  diagrammatically

\begin{align}\label{eq:4e}
\begin{tikzpicture}[>=stealth']
  \coordinate[dot,label={$x_2$}](x2);
\coordinate[dot,below=2.7cm of x2,label=below:{$x_i$}](xi);
  \coordinate[dot,below left=1cm of x2](x1);
  \coordinate[below=.2cm of x1](x1ext){};
  \coordinate[dot,below right=1cm of x2](x3);
  \coordinate[below=.2cm of x3](x3ext){};
  \coordinate[dot,above left=1cm of xi](xip1);
  \coordinate[above=.2cm of xip1](xip1ext){};
  \coordinate[dot,above right=1cm of xi](xim1);
  \coordinate[above=.2cm of xim1](xim1ext){};
  \node[right=.3cm of x1,inner sep=-2pt](k1){$k_1$};
  \node[right=.3cm of xip1,inner sep=-2pt](k2){$k_2$};
  \node[below=0cm of x1]{\vdots};
  \node[below=0cm of x3]{\vdots};
  \draw[very thick,dashed](x2) --node[right]{} (xi);
  \draw[double] (x1)--node[auto]{$p_1$} (x2)--node[auto]{$p_2$} (x3);
  \draw[double] (xim1)--node[auto]{$p_{i-1}$} (xi)--node[auto]{$p_i$}
  (xip1);
  \draw[double] (x1)--(x1ext);
  \draw[double] (x3)--(x3ext);
  \draw[double] (xim1)--(xim1ext);
  \draw[double] (xip1)--(xip1ext);
  \draw[photon] (k1) -- (x2);
  \coordinate[below=.4cm of x2](ang1);
   \coordinate (ang2) at ($(x2)!0.4!(k1)$);
   \begin{pgfonlayer}{background}
     \fill[red] (x2.center)--(ang1.center)--(ang2.center);
   \end{pgfonlayer}
  \draw[photon] (k2) -- (xi);
  \node[below right=.4cm of x3]{$+\quad {\rm res}\ I \quad + $};
  \begin{scope}[xshift=5cm]
  \coordinate[dot,label={$x_2$}](x2);
\coordinate[dot,below=2.7cm of x2,label=below:{$x_i$}](xi);
  \coordinate[dot,below left=1cm of x2](x1);
  \coordinate[below=.2cm of x1](x1ext){};
  \coordinate[dot,below right=1cm of x2](x3);
  \coordinate[below=.2cm of x3](x3ext){};
  \coordinate[dot,above left=1cm of xi](xip1);
  \coordinate[above=.2cm of xip1](xip1ext){};
  \coordinate[dot,above right=1cm of xi](xim1);
  \coordinate[above=.2cm of xim1](xim1ext){};
  \node[right=.4cm of xip1,inner sep=-2pt](k1){$k_1$};
  \node[right=.1cm of xip1,inner sep=-2pt](k2){$k_2$};
  \node[below=0cm of x1]{\vdots};
  \node[below=0cm of x3]{\vdots};
  \draw[very thick,dashed](x2) --node[right]{} (xi);
  \draw[double] (x1)--node[auto]{$p_1$} (x2)--node[auto]{$p_2$} (x3);
  \draw[double] (xim1)--node[auto]{$p_{i-1}$} (xi)--node[auto]{$p_i$}
  (xip1);
  \draw[double] (x1)--(x1ext);
  \draw[double] (x3)--(x3ext);
  \draw[double] (xim1)--(xim1ext);
  \draw[double] (xip1)--(xip1ext);
  \draw[photon] (k1) -- (xi);
  \coordinate[above=.5cm of xi](ang1);
   \coordinate (ang2) at ($(xi)!0.6!(k1)$);
   \begin{pgfonlayer}{background}
     \fill[red] (xi.center)--(ang1.center)--(ang2.center);
   \end{pgfonlayer}
  \draw[photon] (k2) -- (xi);
  \node[below right=.4cm of x3]{$=0$};
\end{scope}
\end{tikzpicture}
\end{align}

Also note that since we are simply adding zero to the sum of Feynman
diagrams, to obtain a manifestly gauge invariant integrand, this means
the original sum of Feynman diagrams, as integrals in Minkowski space
is gauge invariant as expected, despite appearances. 
Since the complex spurious poles cancels in the gauge 
invariant integrand, we are now allowed
to Wick rotate it
and perform the Fourier transform  in Euclidean space. At
this point we use 
the simple Euclidean integration procedure given by Eq. \p{412},
essentially replacing $q$ everywhere in the integrand with the  corresponding term
multiplying $q$ in the exponent. 
Amusingly  after doing this the term $I$ is no longer vanishing, but
gives a non-vanishing result and furthermore, as we explain  
in the following section, it is identified directly to the corresponding
edge terms of the dual diagram.

As a final comment we emphasise that the addition of $I$ can also be
viewed 
as simply a neat trick done in order to perform the Fourier transform
of the $W_{n,m}^{(1,0)}$ sector: the sector is gauge invariant, and in
principle the Fourier transform could be performed directly in
Minkowski space. However by adding  $I$ and thus
removing spurious poles at the level of the integrand we are able to
Wick rotate and make use of the simple Euclidean Fourier
transform \p{412}. The final result should be the same of course whichever
method is used to compute it.

\subsection{Summary of the duality mechanism}
\label{sec:duality-nmhv-wlff}

We are now in a position to pull everything together and prove the
duality   $W_{n,m}^{(1,0)}=W_{m,n}^{(0,1)}$, Eq. \p{W-hat}. 
Having shown the $\xi$-independence of 
$W_{n,m}^{(1,0)}$ (by including integrals similar to the $I$ in~\eqref{eq:6}
above, obtained from edge diagrams of  $W_{m,n}^{(0,1)}$) we can
now Wick rotate all the contributing diagrams (since there are no
remaining 
complex poles to obstruct this) and perform the Euclidean Fourier
transform according to Eq. \p{412}.  We then recognise that cusp diagrams
are directly equivalent  to their graph duals, just as for the MHV sector (e.g.
\eqref{eq:4} equates to \eqref{eq:3} after multiplication of the
appropriate Parke-Taylor factors). Then there are edge diagrams of the
$W_{n,m}^{(1,0)}$ sector (such as \eqref{eq:4b}) which have no dual  in the
$W_{m,n}^{(0,1)}$ sector, but which in fact give zero after performing the
Euclidean Fourier transform. Thus the duality still holds for these
diagrams.  Finally
there are non-vanishing  edge diagrams of the $W_{m,n}^{(0,1)}$
sector, such as~\eqref{eq:3b}. Although these have no dual diagrams in
the $W_{n,m}^{(1,0)}$ 
sector, we find it necessary to add equivalent terms to this sector
(e.g. the expression $I$ in~\eqref{eq:6})  in order to
ensure the absence of spurious poles at the level of the
integrand. As explained above this addition does not affect the Wilson loop form factor
since the integral  vanishes in Minkowski space.
% where our QFT is initially defined. 
However it is crucial for ensuring we have an integrand
without spurious (complex) poles and  hence can safely Wick
rotate. Then we find that the Euclidean space integrals, $I$,  are
equal (up to multiplication by the appropriate Parke-Taylor factors) to
the dual edge diagrams contributing to $W_{m,n}^{(0,1)}$.

In summary then at the Grassmann level  $\kappa + \sigma = 1$  the duality works as follows:

\begin{center}
  \begin{tabular}{|ccc|}
\hline
    $W^{(1,0)}$ sector &&$W^{(0,1)}$ sector
    \\
     (Minkowski) & $\stackrel{\rm Wick}{\longleftrightarrow}$ & (Euclid)  \\
    \hline  cusp diagrams&$\longleftrightarrow$ &cusp diagrams\\[5pt]
    edge diagrams &$\longleftrightarrow$ & 0\\[5pt]
    added terms $I=0$ &$\longleftrightarrow$ & edge diagrams\\
        \hline
  \end{tabular}
\end{center}

\section{Concluding remarks}

In this paper we have given the proof of the new duality for Wilson loop form factors at the first non-trivial NMHV-like level and in the Born approximation. Can we go beyond?

Consider  the general duality \p{duality-gen} in the Born approximation. In this case the cusp diagrams involve several propagators (see Figure \ref{fig:NMHV}) and the corresponding edge diagrams also have a more complicated structure. In particular we need diagrams involving higher-order edge terms in the expansion of the Wilson lines $E_{i+1,i}$, Eq. \p{2.18}. Also, we encounter diagrams of the mixed type, with cups-to-cusp and cusp-to-edge propagators. Nevertheless the mechanism of spurious pole cancellation is expected to be essentially the same. 

We can start with the cusp diagrams for which the duality is evident since it is a duality of planar graphs. These diagrams provide  the physical poles corresponding to vanishing invariant masses, $(k_i+\dots+k_{j-1})^2=y^2_{ij}=0$, or to the distance between two distant points of the Wilson loop contour becoming lightlike, $x_{ij}^2=0$. However they contain various complex spurious poles. These poles are removed by adding the appropriate mixed and edge diagrams. For each spurious pole there are correction terms obtained by sliding an external leg along a propagator. The mechanism is expected to work iteratively, first removing the poles of the pure cusp diagrams, then of the mixed, etc. 

%The duality is evident for cusp diagrams since it is a duality of planar graphs. %. 
%The role of the edge diagrams is to complete the sum of cusp diagrams to gauge-invariant expressions independent of the reference spinor $\xi$. Moreover they also assure the correct singularity structure of $W_{n,m}^{(\kappa,\sigma)}$. This completion of a rational function with known physical poles  is unique. Consequently, applying the duality transformation to it we obtain an expression which has to coincide with the completion for the cusp diagrams in the dual sector $W_{n,m}^{(\kappa,\sigma)}$. So the duality transformation equally relates the sums of edge diagrams in the two sectors. Since the duality holds for both cusp and wedge diagrams, then it holds for their sums, i.e. $W_{n,m}^{(\kappa,\sigma)} \leftrightarrow W_{m,n}^{(\sigma,\kappa)}$.

\begin{figure}[t] 
\hspace*{-3mm}
\psfrag{x0}[cc][cc]{$x_0$}
 \psfrag{x1}[cc][cc]{$x_1$}\psfrag{x2}[cc][cc]{$x_2$}\psfrag{x3}[cc][cc]{$x_3$}\psfrag{x4}[cc][cc]{$x_4$}
 \psfrag{x5}[cc][cc]{$x_5$}\psfrag{x6}[cc][cc]{$x_6$}\psfrag{y0}[cc][cc]{$y_0$}
  \psfrag{y1}[cc][cc]{$y_1$}\psfrag{y2}[cc][cc]{$y_2$}\psfrag{y3}[cc][cc]{$y_3$}\psfrag{y4}[cc][cc]{$y_4$}
 \psfrag{y5}[cc][cc]{$y_5$}  \psfrag{y6}[cc][cc]{$y_6$} 
\includegraphics[width=1.05\textwidth]{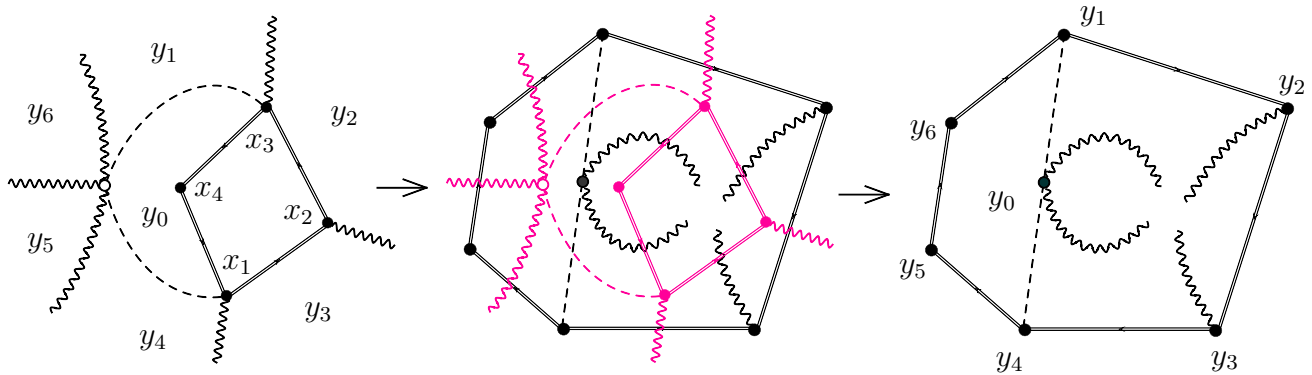}
\caption{Diagrammatic representation of the duality relation $W^{(0,0)}_{4,6} \leftrightarrow W^{(0,0)}_{6,4}$ 
in the one-loop approximation.}
\label{fig:1loop}
\end{figure}

We can also think of the duality beyond the Born approximation. The loop corrections to the vacuum expectation value of  the Wilson loop create UV-divergences.
At loop level the scattering amplitude suffers from IR-divergences. Since the 
Wilson loop form factor is a hybrid observable interpolating between the two,
its perturbative corrections are both UV- and IR-divergent.\footnote{Notice that the divergent part of the Wilson loop form factor automatically satisfies the duality relation (\ref{duality}).
Namely, the IR divergencies of $W_{n,m}$ match the UV divergences of $W_{m,n}$ and vice versa.}
So one needs to introduce a regularisation which can handle both types of divergences. 
%It would be extremely interesting to test whether the loop corrections respect the duality relation \p{duality}.
Instead, we can consider the duality for the {four-dimensional} loop integrands corresponding to Lagrangian insertions into the Born-level object. 

In the planar limit the loop integrands are unambiguously well-defined rational functions. 
So it is natural to expect that the duality works for them similarly to the Born approximation.
Indeed, using the  effective Feynman rules of
Appendix~\ref{sec:effective-rules} together with the Euclidean
Fourier integration rules%
\footnote{We cannot fully justify applicability of the Fourier transform in Euclidean space until we have checked for integrand level cancellation
of spurious poles, but we assume this here.}
 of Appendix~\ref{aC} one can see that the cusp diagrams are dual to
 each other as loop integrands. The corresponding edge diagrams play an auxiliary role cancelling spurious complex poles.
The duality \p{duality} is again translated into a planar graph duality. 
In Figure \ref{fig:1loop} we give an example of the duality in the MHV$\times$MHV sector in the one-loop approximation.  
There the Wilson loop contour is purely bosonic and the scattered
particles are (+1) helicity gluons (this is equivalent to explicitly
performing the
integration over the superspace variable related to  point $y_0$ which the
effective rules naturally give us).
In the left diagram we introduce the region momenta $y_0, y_1 , \ldots , y_6$ associated with faces 
and represent the momentum space integral as an integration over $y_0$.
Multiplying it by the Parke-Taylor prefactor we obtain the contribution to $W_{4,6}^{(0,0)}$,
\begin{align}
  &\int d^4 y_0\frac{
  e^{ix_{23}y_2}e^{ix_{12}y_3}e^{iy_{0}x_{31}} [\xi|y_{10}y_{04}|\xi]^3}
{y_{10}^2 y_{04}^2 [\xi|y_{04}\ket{k_4}\bra{k_6}y_{10}|\xi]
  \bra{p_3}y_{10}|\xi][\xi|y_{10}\ket{k_1}\bra{k_3}y_{04}|\xi][\xi|y_{04}\ket{p_4}}\notag\\
 &\qquad \qquad \times
\frac{\vev{p_4p_1}\vev{p_1 p_2}\vev{p_2 p_3}\vev{k_6k_1}\vev{k_1k_2}\vev{k_2k_3}\vev{k_3k_4}}{\vev{k_1p_2}\vev{p_2k_2}\vev{k_2p_1}\vev{p_1
  k_3}}\,. \label{1loopint}
\end{align}
In the right diagram we use the Euclidean Fourier transform to write
it down immediately in coordinate space
and integrate over position $y_0$ of the interaction vertex. Its contribution to $W_{6,4}^{(0,0)}$ coincides with \p{1loopint}.
So we see the duality at the level of the integrand.

{There are several directions for further investigations.} It is well known that the Born-level amplitudes have a remarkable dual superconformal symmetry which, combined with the native superconformal symmetry, results in a Yangian structure  \cite{Drummond:2008vq,Berkovits:2008ic,Beisert:2008iq,Drummond:2009fd}. As a result, the form of the amplitude is completely determined by this powerful symmetry and the requirement of absence of spurious poles.  In this context we may ask the question if the new duality found in this paper could be a  manifestation of some hidden symmetry? The first step in this direction should be to elucidate the role of conformal symmetry. It is supposed to simultaneously act on the Wilson loop component of the form factor as a local symmetry, and on its amplitude component as a non-local symmetry. This issue is under investigation.  

{It would also be interesting to understand how to properly regularise the loop correction integrals so that the duality still holds at loop level. 
Another challenging problem is to find a strong coupling or AdS/CFT analog of this duality. }
 
\section*{Acknowledgements}

We profited from numerous discussions with {Simon Caron-Huot and \"Omer G\"urdo\u{g}an}.
G.K. would like to thank  \"Omer G\"urdo\u{g}an for collaboration at the early stage of this project.
  We acknowledge partial support by the French National Agency for
  Research (ANR) under contract StrongInt (BLANC-SIMI-4-2011). The
  work of D.C. has been   partially supported by the RFBR grant
  14-01-00341. The work of P.H. has been partially supported by an STFC
  Consolidated Grant ST/L000407/1. P.H. would also like to thank the 
  CNRS for financial support and LAPTh for hospitality  where part of this work was done.

\appendix

\section{Chiral lightlike Wilson loop in LHC superspace}\label{apA}

\subsection{Chiral lightlike Wilson loop  }

The conventional formulation of a {\it chiral} supersymmetric Wilson line on a segment of the line in $(x,\theta)-$superspace 
\begin{align}\label{1.3}
x_{\alpha\dot\alpha}(t) = x_{\alpha\dot\alpha} - t\,  p_\alpha \tilde p_{\dot\alpha}\,, \qquad \q_\alpha^A(t) = \q_{\alpha}^A - t\, \omega^A p_\alpha \,, \qquad t\in[0,1]\,,
\end{align} 
has the form  (recall \p{127})
\begin{align}\label{27}
\cE =   P \exp \left\{  -i \int_0^1 dt \Big[\frac1{2} \tilde p^\da  p^\a \cA_{\a\da}(x(t),\q(t)) + \omega^A   p^\a\cA_{\a A}(x(t),\q(t)) \Big]  \right\}\ .
\end{align}
The corresponding covariant derivatives $\cD = \pa + \cA$ transform under a gauge group with ordinary, harmonic-independent parameters,
\begin{align}\label{28}
\cD \ \to \   e^{\tau(x,\q)}\ \cD\  e^{-\tau(x,\q)} \,.
\end{align}
Consequently the Wilson line transforms as follows:
\begin{align}\label{29}
 \cE \ \to \   e^{\tau(x(1), \q(1))}\ \cE\  e^{-\tau(x(0),\q(0))} \,.
\end{align}
A complete gauge-invariant lightlike Wilson loop is obtained by gluing together $n$ consecutive segments and closing the contour (with the identification $n+1 \equiv 1$),
\begin{align}\label{224}
{\cal W}_n = \frac1{N} \tr\prod_{i=1}^n \cE_{i+1,i}\,.
\end{align}

\subsection{Bridge transformation to LHC superspace}

For our purposes we need to express the Wilson loop \p{224} in terms of the unconstrained prepotentials $A^{++}, A^+_\da$ from \p{a1}.  To this end we need to relate the covariant derivatives $\cD$ with the transformation \p{28} to the derivatives $\nabla$ with the transformation  \p{a3}. The key observation is that the harmonic derivative $\pa^{++}$ needs no connection for the gauge group with harmonic-independent parameters $\tau(x,\q)$, $\cD^{++} = \pa^{++}$. This suggests to relate it to $\nabla^{++}$ from \p{a2} by a  {\it generalised gauge transformation},
\begin{align}\label{213}
\pa^{++} = h^{-1} \nabla^{++}  h = h^{-1} (\pa^{++} + A^{++}) h
\end{align}
or equivalently
\begin{align}\label{214}
A^{++}(x,\q^+,u) = - (\pa^{++} h) h^{-1} = h\, \pa^{++} h^{-1}\,.
\end{align}
Here the `parameter' $h(x,\q,u)$ depends on all the LHC superspace variables. It undergoes gauge transformations under both gauge groups, 
\begin{align}\label{211}
h \ \to \  e^{\Lambda(x,\q^+,u)}\ h\ e^{-\tau(x,\q)}\,. 
\end{align}
In the Abelian case this becomes $\delta h(x,\q,u) = \Lambda(x,\q^+,u)-\tau(x,\q) $ and we see that the combined gauge transformations of both types cannot gauge away the entire content of the general superfield $h$. Hence, despite the appearance $h$ is not a pure gauge. In the same way, \p{213} and \p{214} are not gauge transformations but rather field redefinitions. We call the new object $h$ a `gauge bridge' \footnote{The terminology originates from the harmonic superspace formulation of $\cN=2$ SYM \cite{Galperin:1984av}. A similar object exists in the Ward construction for self-dual non-supersymmetric Yang-Mills  \cite{Ward:1977ta,Galperin:1987wc}. }  relating  the $\tau-$frame with harmonic-independent  parameters and the $\Lambda-$frame with analytic parameters. 

Relation \p{214} is a differential equation on the  harmonic sphere $S^2$. In it we consider the chiral-analytic prepotential $A^{++}$ as given and the bridge $h$ as the unknown. This equation has a solution defined up to arbitrary $\tau$ and $\Lambda$ gauge transformations, so the bridge $h$ {\it cannot be obtained unambiguously} from the prepotential. 

Once the bridge has been found, it can be used to convert any gauge covariant object from the $\tau-$frame to the $\Lambda-$frame or vice versa. In particular, all covariant derivatives $\cD$ can be converted to $\nabla$,
\begin{align}\label{}
\nabla = h\, \cD\, h^{-1}
\end{align}
transforming according to \p{a3}. Then we can do the same with the Wilson line \p{27}:
\begin{align}\label{225}
E  = h(x(1),\q(1),u)\ \cE\ h^{-1}(x(0),\q(0),u) 
\end{align}
with the gauge transformation 
\begin{align}\label{226}
E  \ \to \   e^{\Lambda(x(1), \q^+(1),u)}\ E \  e^{-\Lambda(x(0),\q^+(0),u)} \,.
\end{align}
Note that we use the same harmonic variable $u$ all along the segment. 

Conversely, starting from the $\Lambda-$frame Wilson line  and transforming it back to the $\tau-$frame, we obtain an object which does not depend on the harmonics:
\begin{align}\label{229}
\cE = h^{-1}(x(1),\q(1),u)\ E(u)\ h(x(0),\q(0),u)\,, \qquad \pa^{++} \cE =0\,.
\end{align}
Indeed, the harmonic dependence of the $\Lambda-$frame Wilson line \p{225} comes from the bridge $h$. The inverse transformation \p{229} removes the bridge and with it the $u-$dependence.
This property allows us to choose the harmonics $u$ on a given segment  as we like. A judicious choice \cite{Rosly:1996vr,Mason:2010yk} is to identify the harmonic $u^+$ with the chiral spinor defining the direction of the lightlike line,
\begin{align}\label{}
u^+_\a \equiv  p_\a\,,  
\end{align}
and $u^-$ with the $SU(2)$ conjugate $\bar p_\a$. With this choice we obtain (see \p{1.3})
\begin{align}\label{2.15}
E &=   P \exp \left\{  -i \int_0^1 dt \left[\frac1{2}\tilde p^{\da}p^\a A_{\a\da}(x(t),\q(t),\ket{p}) + \omega^{A} p^\a A_{\a A}(x(t),\q(t),\ket{p}) \right]  \right\}\nt
& =  P \exp \left\{  -i \int_0^1 dt \left[\frac1{2}\tilde p^{\da} A^+_{\da}(x(t),\q(t),\ket{p}) + \omega^{A} A^+_{A}(x(t),\q(t),\ket{p}) \right]  \right\} \nt
& =  P \exp \left\{  -\frac{i}{2} \int_0^1 dt \ \tilde p^{\da} A^+_{\da}(x(t),\vev{p\q},\ket{p})  \right\}\,.
\end{align}
In the last relation we have used the $\Lambda-$frame property $A^+_{A}=0$, i.e. $\nabla^+_A=  \pa^+_A $. The gauge connection contributing to the Wilson line is the prepotential $A^+_\da(x,\q^+,u)$ from \p{a1} with $\q^{+A} = u^{+\a}\q^A_\a(t)  = p^\a(\q^A_{\a} - t \omega^A p_\a)= \vev{p\q^A}$. So, the only dependence on the position along the lightlike segment is in the space-time coordinate $x(t)$.

Let us now glue together the different segments of a Wilson loop according to \p{224}. Consider two adjacent segments:
\begin{align}\label{2.34}
 &\cE_{i,i-1} = h^{-1}(x_i,\q_i,  p_{i-1})\ E_{i,i-1} \ h(x_{i-1}, \q_{i-1},  p_{i-1}) \nt
 & \cE_{i+1,i} = h^{-1}(x_{i+1},\q_{i+1},  p_{i})\ E_{i+1,i}\ h(x_{i}, \q_{i},  p_{i}) \,.
\end{align}
In their product $  \cE_{i+1,i} \cE_{i,i-1}$ the bridges at the cusp point $i$ depend on the same superspace coordinates $x_{i}, \q_{i}$ but on different harmonics $p_{i-1}$ and $p_i$. The two adjacent bridges define a new object,
\begin{align}\label{233}
h(x_i,\q_i,  p_{i})\ h^{-1}(x_i,\q_i,  p_{i-1}) := U(x_i,\q_i;  p_{i}, p_{i-1})\,.
\end{align}

This object  has been discussed in detail in \cite{Chicherin:2016qsf}. The bridge $h(x,\q,u)$ relates the harmonic-independent $\tau-$ gauge frame  and the analytic $\Lambda-$gauge frame. The new bridge
\begin{align}\label{Uh}
U(x,\q;u,v) = h(x,\q,u)\ h^{-1}(x,\q,v) 
\end{align}
is inert under the $\tau$-frame gauge transformations but transforms with respect to two  analytic frames  with harmonics $v$ and  $u$ (see \p{211}),
\begin{align} \label{gb}
 U(x,\q;u,v) \ \rightarrow \  e^{\Lambda(x,\q^+,u)}\ U(x,\q;u,v)\ e^{-\Lambda(x,\q^+,v)}\,.
\end{align}
We call this object a {\it bilocal bridge}, where the biloclality refers only to the harmonic variables. 
An explicit expression for $U$ in terms of the prepotential  $A^{++}$ from \p{a1}
can be found by solving the harmonic differential equation that follows from \p{Uh} and from  the definition \p{213} of the bridge $h$,
\begin{align}\label{}
\nabla^{++}_u U(u,v) = 0\,,
\end{align} with the obvious boundary condition $U(u,u)=1$. The solution  \cite{Lovelace:2010ev}  is given in \p{U'}. 

In conclusion,  coming back to the closed Wilson loop \p{224}, we obtain the {\it formulation of the chiral lightlike Wilson loop in LHC superspace} by gluing together Wilson line segments  \p{2.15} and using bilocal bridges \p{Uh} as `glue', Eq.~\p{2.17}, where  $E_{i+1,i}$ is defined in \p{2.18}. Comparing the gauge transformations of the bilocal bridge $U$ \p{gb} and of the Wilson line segments $E$ \p{226}, we clearly see the role  of the bridges: they adjust the transformation properties of the segments at the cusp points.

\subsection{Comments on the twistor formulation of Mason and Skinner}\label{aA1}

The construction of lightlike Wilson loops presented here is similar in spirit to the twistor  formulation of Mason and Skinner in \cite{Mason:2010yk}. After establishing the LHC/twistor dictionary (see \cite{Chicherin:2016fac}), we can make a detailed comparison. 

The noticeable difference concerns the claim  in \cite{Mason:2010yk} that the entire Wilson loop can be written as a product of `parallel propagators' $U$ \cite{Mason:2005zm,Mason:2010yk,Bullimore:2011ni},
\begin{align}\label{MS}
{\cal W}^{\rm M{\&}S}_n &= \frac1{N} \tr\prod_{i=1}^n U(x_i,\q_i;  p_{i}, p_{i-1})\,. %= \tr \prod_{i=1}^n  h(x_i,\q_i, p_{i-1})\ h(x_i,\q_i, p_{i})^{-1} \nt
%& = \tr \prod_{i=1}^n h(x_i,\q_i, p_{i})^{-1}\ h(x_{i+1},\q_{i+1}, p_{i}) \,. %:=  \tr \prod_{i=1}^n {\cal U}(x_i,\q_i;x_{i+1},\q_{i+1}; p_{i})  \,.
\end{align}
Equation (6.16) (see also eq.~(2.17))  in \cite{Mason:2010yk} displays this form of the Wilson loop. Comparing with our formulation \p{2.17}, we see that the Wilson line edge factors $E_{i+1,i}$ are missing in \p{MS}. One would wonder how such an incomplete expression can be gauge invariant? Indeed, from \p{gb} we see that the product $U(x_{i+1},\q_{i+1},  p_{i+1},  p_{i}) U(x_i,\q_i,  p_{i},  p_{i-1}) $ is not gauge invariant at point $i$; the role of the edge factor $E_{i+1,i}$ is precisely to adjust the transformation properties. 

The attentive reader may notice that on each segment of the Wilson loop Mason and Skinner impose the  condition $\tilde p^\da  A^+_{\da}=0$ (in our notation; see the text before their equation (6.14)). This condition would indeed trivialise \p{2.18} but it is not clear where it comes from. It looks like a `floating gauge condition', i.e.  the light-cone gauge $\xi^{\da} A^+_\da=0$ \p{a8} but with $\xi  \equiv \tilde p_i$ for each segment. In our understanding, if we change the gauge-fixing condition from one segment of the Wilson loop to the other, we would break gauge invariance. The edge  contribution $E_{i+1,i}$ in \p{2.17} is an indispensable part  of the definition of the Wilson loop. In Section~\ref{sec:nmhv} we have shown explicit examples where the edges contribute essential pieces of the complete  NMHV Wilson loops. In fact, they are responsible for obtaining a gauge-invariant, i.e. $\xi-$independent result.

\section{Feynman rules}\label{apB}

\subsection{Propagators and on-shell states}

To quantise the theory we use the light-cone (or axial or CSW) gauge \cite{Cachazo:2004kj,Boels:2006ir}
\begin{align}\label{a8}
\xi^\da A^+_\da=0\,,
\end{align}
where the gauge-fixing parameter $\xi^\da$ is an antichiral commuting spinor (``reference spinor"). This gauge has the advantage that the cubic  Chern-Simons term in \p{CS} vanishes, so the  interaction comes only from  $L_Z$ \p{lint}. However, it introduces specific spurious poles in the propagators which require careful treatment. We find that the correct way is to implement all Feynman rules in momentum space, even though the Wilson loop is defined in position space. The Fourier integrals from position to momentum space are initially defined in Minkowski space. Only after we have made sure that the sum of all Feynman diagrams is free from spurious poles, we can Wick rotate to Euclidean space and evaluate the Fourier integrals by the simple formulas in Appendix   \ref{aC}.

The Feynman rules in the gauge \p{a8} have been worked out in \cite{Chicherin:2016fac} in position space and their momentum space equivalents in  \cite{Chicherin:2016qsf}.  Here we summarise the effective rules after  the Grassmann and harmonic integrations at the vertices have been carried out.

The propagators are determined from the (non-diagonal) quadratic part of the Chern-Simons Lagrangian \p{CS}:\footnote{The bilinear term in $L_Z$ is treated as a bivalent vertex.}
\begin{align}
&\vev{A^{++}(q,\q^+_1,u_1) A^{++}(-q,\q^+_2,u_2)}= 4\pi \delta^2\left( \bra{u^+_1} q |\xi] \right) \delta (u_1, u_2)\, \delta^{4}\left(\vev{ u^+_1 \q_{12} }\right)\,, 
\label{prop1}\\
&\langle A^{+}_{\da}(q,\q^+,u_1) A^{++}(-q,0,u_2) \rangle=
\frac{4\ti  \xi_{\da}}{\bra{u^+_1}q | \xi]}  \,
\delta(u_1,u_2) \,\delta^{4}\left(\vev{u^+_1 \q_{12}}\right)\,,  \label{prop2} 
\end{align}
and $\vev{A^+_{\da} A^+_\db}=0$. The harmonics are auxiliary variables which are integrated out in gauge-invariant quantities.
The harmonic integrations originate from the interaction vertices in the Lagrangian $L_{Z}$ \p{lint}
and in the bilocal bridge $U$  \p{U'}. They are implemented with the help of delta functions. Firstly, the harmonic delta-function $\delta(u_1,u_2)$ 
identifies $u_1^{\pm} = u_2^{\pm}$, so each line effectively carries only one harmonic. Secondly, the remaining harmonic of the propagator 
$\vev{ A^{++} A^{++}}$ is integrated out by means of the delta-function $\delta^2$, Eq. \p{prop1}, resulting in 
the substitution $u^+ \to q|\xi]$ and the Jacobian factor $1/q^2$, i.e.
\begin{align} \label{harmint}
\int d u \, \delta^2 \left( \bra{u^+} q |\xi] \right) {\cal R}(u) = \frac{1}{\pi\, q^2} {\cal R}( q|\xi])\,,
\end{align} 
for a homogeneous function of degree zero ${\cal R}(u)$. In the effective Feynman rules all harmonic integrations are already implemented.

Another effective Feynman rule concerns the emission of $n$ external states and of one propagator from an interaction vertex in the Lagrangian $L_Z$.   Denoting by $K=\sum_1^n k_i$ and $Q_K = \sum_1^n \ket{k_i}\eta_i$ the total (super)momentum of the $n$  particles and using the Feynman rules \p{prop1} we find:
\begin{align}
&\int d^4 x_0  d^8 \theta_0\,  \left\langle (k_1,\eta_1)  \ldots (k_n,\eta_n) | L_{Z}(x_0,\theta_0) \,   A^{++}(x,\theta^+,u) \right\ket{0} \notag\\
&= \int d^4x_0  d^8 \q_0 \,  e^{iKx_0+Q_K\q_0}\int \frac{d^4 q}{(2\pi)^4}  \frac{e^{iq(x_0-x)}}{q^2}  \delta^4\left( \bra{\q_0-\q} q |\xi] \right)\ \frac{1}{[\xi|q\ket{k_1}\vev{k_1 k_2}\ldots \bra{k_n} q |\xi]}\nt
%&= \frac{e^{i K x}}{K^2}\, \frac{1}{[\xi|K\ket{k_1}\vev{k_1 k_2}\bra{k_2} K |\xi]}\ \int d^8 \q_0 \,  e^{Q_K\q_0} \delta^4\left( \bra{(\q_0-\q)} K |\xi] \right) \nt
%\nt
&= \frac{e^{i K x + Q_K \theta}}{K^2}\, \frac{\delta^4\left( \bra{Q_K} K |\xi] \right) }{[\xi|K\ket{k_1}\vev{k_1 k_2}\ldots \bra{k_n} K |\xi]}\,. 
\end{align}

The prepotential $A^{++}$ plays the leading role. The diagrams with propagators  $\vev{A^{++} A^{++}}$  
provide the main contribution (cusp diagrams). The prepotential $A^{+}_{\da}$ 
and the propagator $\vev{A^{++} A^+_{\da}}$ appear in edge diagrams which are needed to restore gauge invariance (i.e., to eliminate the reference spinor $\xi$ and its spurious  poles).
The only dependence on $A^{+}_{\da}$ comes from the edges $E_{i +1,i}$ of the Wilson loop. The prepotential $A^+_\da$ slides along an edge of the Wilson loop contour, see \p{2.18}. 
At  NMHV level we deal with the linear approximation of the Pexp.
The corresponding line integral concerns only the exponential factors (Fourier or wave function), e.g. 
$\int_0^1 dt \, e^{iq (x_i - t p_i)}= \frac{1}{i(q p_i)}\left( e^{i q x_{i}} - e^{i q x_{i+1}} \right)$ . 
 
In the following we deal mostly with harmonics $u^+$ carrying $U(1)$-charge $(+1)$, so we omit the $+$ index of the harmonics 
and the Grassmann variables for the sake of brevity.

\subsection{Effective Feynman rules}
\label{sec:effective-rules}

To obtain effective Feynman rules we carry out all  harmonic and Grassmann  integrations.

\subsubsection{Vertices}

The bilocal bridge $U(x,\q;{p_1},{p_2})$, Eq. \p{U'}, at the lowest order in the coupling  equals $1$:

\begin{align}\label{eq:2}
\parbox[c]{0.3\textwidth}{\hspace*{-20mm}
 \psfrag{p1}[cc][cc]{$p_2$}\psfrag{p2}[cc][cc]{$p_1$} 
  \psfrag{u1}[cc][cc]{$u_1$}\psfrag{u2}[cc][cc]{$u_2$}\psfrag{u3}[cc][cc]{$u_3$}\psfrag{un}[cc][cc]{$u_n$}
  \psfrag{eq1}[lc][cc]{$\displaystyle =1$}
  \psfrag{eq2}[lc][cc]{$\displaystyle =\frac{\vev{p_1p_2}}{\vev{p_1 u_1}\vev{u_1 u_2}\dots  
   \vev{u_n p_2}}$}
    \psfrag{eq3}[lc][cc]{$\displaystyle =\frac{1}{\vev{u_1 u_2}\vev{u_2 u_3}\dots  
   \vev{u_n u_1}}$}
 \includegraphics[width=0.3\textwidth]{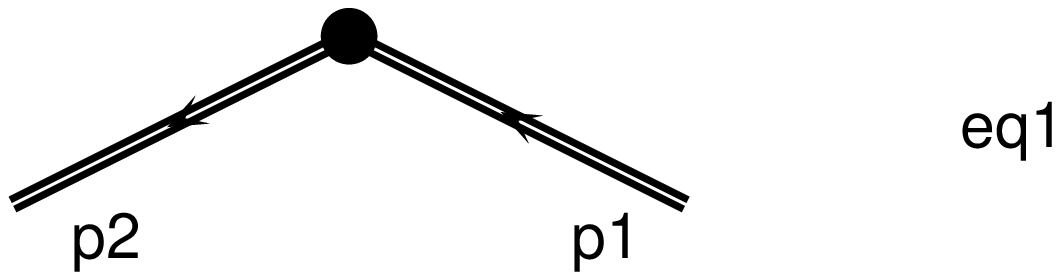}}
\end{align}
\vskip5mm
Higher-order vertices in the expansion of $U(x,\q;{p_2},{p_1})$. 
In this example the line  with $u_2$ ends on a superstate $(k,\eta)$
and depicts particle emission. The harmonic is identified with the particle momentum, $u_{2} \to \ket{k}$. The other lines are internal. If an internal line depicts the  propagator 
$\vev{A^{++}(q,u)A^{++}(-q,u)}$, its harmonic is substituted by $u^+ \to q|\xi]$, see eq.~\p{harmint}.
If an internal line depicts the propagator $\vev{A^{++} A^{+}_{\da}}$, then $u^+$ is identified with the lightlike direction spinor $\ket{p}$
of the edge where the prepotential $A^+_{\da}$ lives.

\begin{align} \label{B7}
\parbox[c]{0.3\textwidth}{\hspace*{-20mm}
 \psfrag{p1}[cc][cc]{$p_2$}\psfrag{p2}[cc][cc]{$p_1$} 
  \psfrag{u1}[cc][cc]{$u_1$}\psfrag{u2}[cc][cc]{$u_2$}\psfrag{u3}[cc][cc]{$u_3$}\psfrag{un}[cc][cc]{$u_n$}
  \psfrag{eq1}[lc][cc]{$\displaystyle =1$}
  \psfrag{eq2}[lc][cc]{$\displaystyle =\frac{\vev{p_1p_2}}{\vev{p_1 u_1}\vev{u_1 u_2}\dots  
   \vev{u_n p_2}}$}
    \psfrag{eq3}[lc][cc]{$\displaystyle =\frac{1}{\vev{u_1 u_2}\vev{u_2 u_3}\dots  
   \vev{u_n u_1}}$}
 \includegraphics[width=0.3\textwidth]{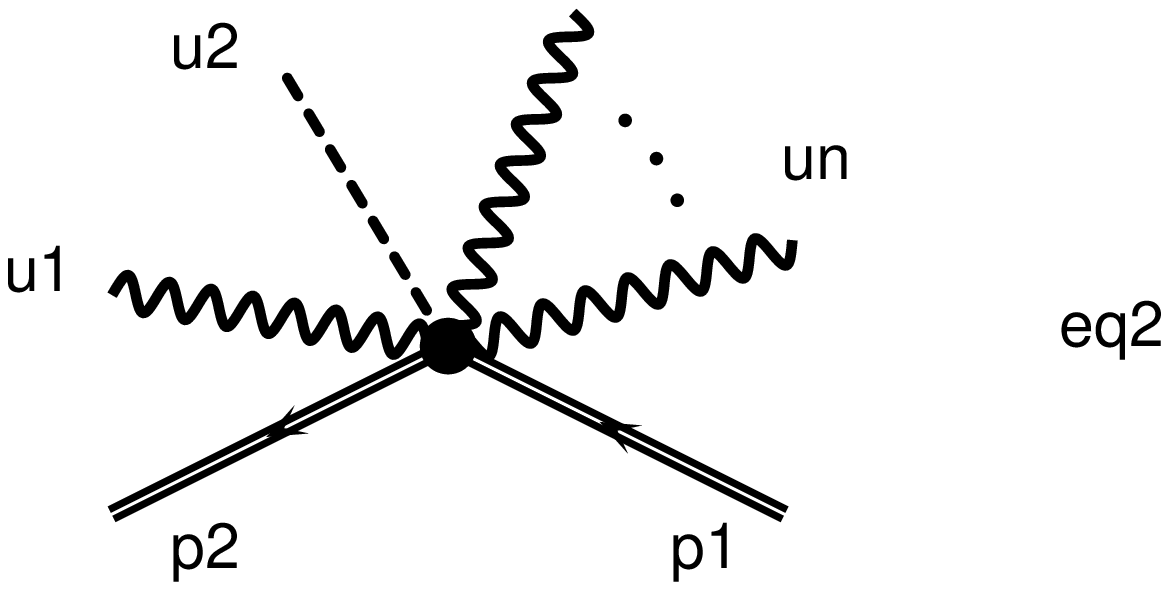}}
\end{align}

\vskip5mm
Interaction vertex $L_{Z}$, eq.~\p{lint}. Lines  $i = 1,3$ correspond to particle emission. The other lines  are internal. The harmonics are identified as above.

\begin{align} \label{B8}
\parbox[c]{0.3\textwidth}{\hspace*{-20mm}
 \psfrag{p1}[cc][cc]{$p_2$}\psfrag{p2}[cc][cc]{$p_1$} 
  \psfrag{u1}[cc][cc]{$u_1$}\psfrag{u2}[cc][cc]{$u_2$}\psfrag{u3}[cc][cc]{$u_3$}\psfrag{un}[cc][cc]{$u_n$}
  \psfrag{eq1}[lc][cc]{$\displaystyle =1$}
  \psfrag{eq2}[lc][cc]{$\displaystyle =\frac{\vev{p_1p_2}}{\vev{p_1 u_1}\vev{u_1 u_2}\dots  
   \vev{u_n p_2}}$}
    \psfrag{eq3}[lc][cc]{$\displaystyle =\frac{1}{\vev{u_1 u_2}\vev{u_2 u_3}\dots  
   \vev{u_n u_1}}$}
 \includegraphics[width=0.3\textwidth]{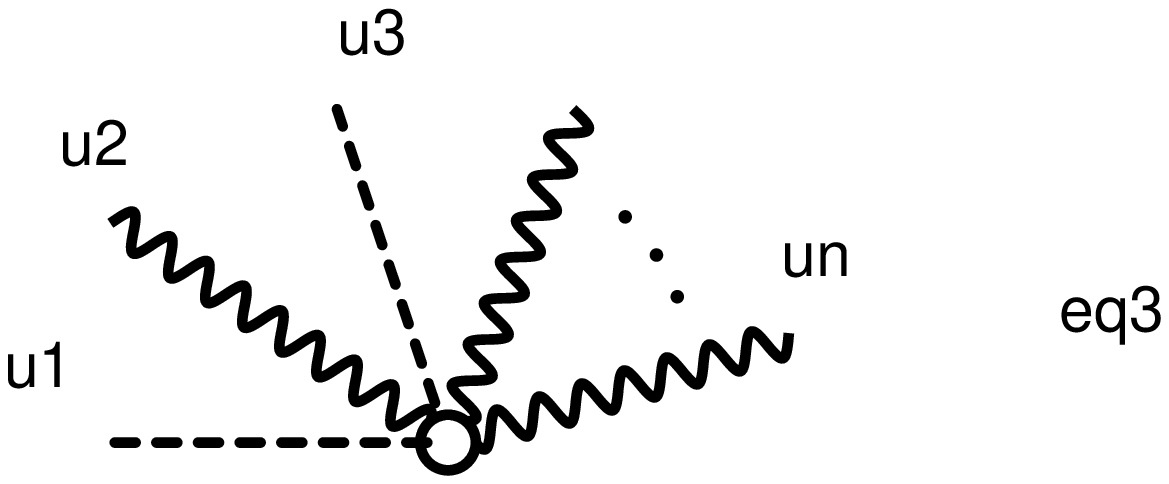}}
\end{align}

\subsubsection{Propagators}

\begin{enumerate}  

\item Propagator $\vev{ A^{++}A^{++}}$ between two Wilson loop cusps. The harmonic of the vertices \p{B7} and \p{B8} where this  propagator ends is  replaced by $u^+ \to q|\xi]$.

 \begin{align}\label{B9}
 \begin{tikzpicture}
     \node[dot,label={$(x_1,\theta_1,q|\xi])$}](a){};
     \node[right=3cm of a,dot,label={$(x_2,\theta_2,q|\xi])$}](b){};
     \draw[very thick,dashed]    (a) -- (b);
 \end{tikzpicture}
&= \int \frac{d^4 q}{4\pi^2} \,e^{iqx_{12}} \frac1{q^2} \delta^4\big( \bra{ \theta_{12} }
     q |\xi]  \big)
 \end{align}

\item  Emission of a superparticle with (super)momentum $(k,Q_k\equiv \eta \ket{k})$ from a cusp of the Wilson loop. The  wave function $\vev{ k,Q_k| A^{++}(x,\q,\ket{k})|0} $ is   

\begin{align}\label{B10}
\begin{tikzpicture}
     \node[] (a){};
     \node[right=3cm of a,dot,label={$\qquad \big(x,\q,\ket{k}\big)$}]{}(b);
     \draw[photon]  (a) --  node[above]{$(k,Q_k,|k\rangle)$} (b);
 \end{tikzpicture}
&\ =\ e^{ikx+Q_k \q}
\end{align}

\item Emission of a superparticle from an interaction vertex $L_{Z}$.

\begin{align}\label{B11}
\begin{tikzpicture}
     \node[] (a) {};
     \node[right=3cm of a,dotw](b){};
     \draw [photon]    (a) -- node[above]{$(k,Q_k,\ket{k})$} (b);
 \end{tikzpicture}
&=1
\end{align}

\item Propagator $\vev{ A^{++}A^{++}}$ between  an interaction vertex and a cusp $(x,\theta)$.
The momentum $K$ and supercharge $Q_K$ flow into this line
from the side of the vertex.    
     
  \begin{align}\label{B12}
 \begin{tikzpicture}
     \node[dotw] (a){};
     \node[right=3cm of a,dot,label={$\qquad \quad (x,\theta,K|\xi])$}](b){};
     \draw[very thick,dashed] (a) -- node[above]{$(K,Q_K,K|\xi])$} (b);
 \end{tikzpicture}
&=    \frac1{K^2} e^{iKx+ Q_K\q}   \delta^4\big(\bra{Q_K}  K |\xi]\big)
\end{align}

\item  Interaction vertex connected with a Wilson line segment $W_{12}$ by a propagator $A^{++} A^+_\da$ sliding along the edge parametrised by $x(t) = x_1 - t \ket{p}[p|$, $p \equiv x_1 - x_2$, $t\in[0,1]$. The (super)momentum $(K,Q_K)$ flows into this line
from the side of the vertex.

 \begin{align}\label{B13}
\begin{tikzpicture}[>=stealth']
    \coordinate[dotw] (a);
    \coordinate[right=3cm of a](b);
    \coordinate[above=.8cm of b,dot,label={$(x_2,\theta_2,\ket{p})$}](c);
     \coordinate[below=.8cm of b,dot,label=below:{$(x_1,\theta_1,\ket{p})$}](d);
     \draw[->,very thick,dashed]         (a) -- node[above]{$(K,Q_K,\ket{p})$} (b);
     \draw[double]         (c) --  (d);
 \end{tikzpicture}
&
 \raisebox{1.5cm} { $\displaystyle
  = \frac{[p\xi]e^{
  Q_K \theta_1}}{\bra{p} K|\xi]\,\bra{p} K|p]}\delta^4\big(\vev{Q_K p}\big)\big(e^{iKx_1}-e^{iKx_2}\big)$ }
 \end{align} 

\item Cusp $(x_0,\theta_0)$ connected with $W_{12}$ by a propagator sliding along the edge.
Note that $\vev{p \theta_{10}} = \vev{p \theta_{20}}$.   
  
 \begin{align}\label{B14}
\begin{tikzpicture}[>=stealth']
    \coordinate[dot,label=above:{$(x_0,\theta_0,\ket{p})$}] (a);
    \coordinate[right=3cm of a](b);
    \coordinate[above=.8cm of b,dot,label={$(x_2,\theta_2,\ket{p})$}](c);
     \coordinate[below=.8cm of b,dot,label=below:{$(x_1,\theta_1,\ket{p})$}](d);
     \draw[->,very thick,dashed]         (a) -- (b);
     \draw[double]         (c) --  (d);
 \end{tikzpicture}
&
 \raisebox{1.5cm} { $\displaystyle = \int \frac{d^4q}{4\pi^2} \frac{[p\xi]}{\bra{p} q|\xi] \bra{p} q|p]}\delta^4\big(\vev{\theta_{01} p}\big)\big(e^{iqx_{01}}-e^{iqx_{02}}\big)$ }
   %\end{array}
 \end{align}
 
\end{enumerate}

 \section{Fourier transforms} \label{aC}
 
 In this paper we encounter Fourier integrals  of the type 
\begin{align}\label{}
\int d^4 q\  \frac{ e^{i q x}}{q^2} \cR([\xi|q)  \,,
\end{align} 
where  $\cR([\xi|q)$ is a homogeneous function  of degree zero  of the spinor $[\xi|q$. The way this integral is computed heavily depends on the space-time signature. In Euclidean space we can use the following trick. First, we insert a harmonic integral over $S^2$ and write
\begin{align}
\int d^4 q\  \frac{ e^{i q x}}{q^2} \cR([\xi|q) &= \pi\int d^4 q \ e^{i q x}\, \int du\ \delta^2([\xi|q\ket{u^+})\, \cR(u^+) \,.
\end{align}
The role of the complex delta function is to identify the $SU(2)_L$ harmonic $u^+ \sim q|\xi]$, up to a phase factor which drops out in the homogeneous function $\cR(u^+)$. The factor $1/q^2$ is the corresponding Jacobian (for detail see Appendix A in \cite{Chicherin:2016fac}). Then we decompose the dot product $$2(q \cdot x) = [\bar\xi|q\ket{u^+} [\xi|x\ket{u^-} + [\xi|q\ket{u^-} [\bar\xi|x\ket{u^+} -  [\bar\xi|q\ket{u^-} [\xi|x\ket{u^+} - [\xi|q\ket{u^+} [\bar\xi|x\ket{u^-}$$
in the basis formed by the $SU(2)_L$ harmonics $u^\pm_\a$ and their $SU(2)_R$ analogs $\xi_\da, \bar\xi_\da$ with $\vev{u^+ u^-}=1$ and $[\xi\bar\xi]=1$. Next we swap the harmonic and Fourier integrals and use the delta function to lift the integral over the complex variable $[\xi|q\ket{u^+}$ (and its conjugate $[\bar\xi|q\ket{u^-}$). The remaining Fourier integral over $[\xi|q\ket{u^-}$ produces another delta function, so
\begin{align}\label{}
 \int d^4 q \ e^{i q x}\, \delta^2([\xi|q\ket{u^+}) =4\pi^2\delta^2([\bar\xi|x\ket{u^+}) \,.
\end{align}
Finally, we do the harmonic integral with the help of the new delta function and obtain
\begin{align}\label{412}
\int \frac{d^4 q}{4\pi^2}\  \frac{ e^{i q x}}{q^2} \cR([\xi|q)   = \frac{1}{x^2}  \, \cR([\bar\xi|x)\,.
\end{align}

As an example, consider  the integral 
\begin{align}\label{e330}
\int \frac{d^4 q}{4\pi^2}\  \frac{ e^{i q x}}{q^2}\, \frac{[\xi|q\ket{p_1}}{[\xi|q\ket{p_2}} =  \frac{1}{x^2}  \, \frac{[\bar\xi|x\ket{p_1}}{[\bar\xi|x\ket{p_2}}\,.
\end{align}
Its computation in Euclidean signature is straightforward, as shown above. However, were this integral to be evaluated with Minkowski signature, we would have to deal with the {\it complex pole} at $[\xi|q\ket{p_2}=0$. This is a rather nontrivial task which we need not address. 

Another type of Fourier integral that we encounter is obtained by  combining two integrals of the type \p{e330}:
\begin{align}\label{e331}
\int \frac{d^4 q}{4\pi^2}\, e^{i q x}\ \frac{[\zeta \xi] }{[\xi|q\ket{p_2} [\zeta|q\ket{p_1}}  &= \frac1{\vev{p_1 p_2}}\int \frac{d^4 q}{4\pi^2}\  \frac{ e^{i q x}}{q^2}\, \left[  \frac{[\xi|q\ket{p_1}}{[\xi|q\ket{p_2}} -  \frac{[\zeta|q\ket{p_1}}{[\zeta|q\ket{p_2}}\right] \nt
&    =\frac{[\bar\zeta \bar\xi] \vev{p_1 p_2}}{[\bar\xi|x\ket{p_2} [\bar\zeta|x\ket{p_1}} \,. 
\end{align}
We remark that in Minkowski signature, with the choice $[\zeta| =\tilde p_1$, this integral becomes a {\it contact term} $\sim \delta( x\cdot p_1)$. At the same time, in Euclidean signature it is a {\it rational function}. This example clearly illustrates the drastic difference between the two space-time signatures. It also explains why we prefer to do all our Fourier integrals after the Wick rotation to  Euclidean space.

\section{Cancellation of spurious poles: boundary cases}
\label{sec:boundary-cases}

{Here we discuss in more details the cancellation of spurious poles between cusp and edge diagrams at 
the Grassmann level $\kappa + \sigma = 1$.}

\subsection*{ MHV$\times$NMHV sector}

From the discussion in section~\ref{sec:spur-pole-canc} we see that
spurious poles in the $W^{(0,1)}$ sector are cancelled by nearby diagrams. 
In particular, in~\eqref{eq:4d} we see that spurious
poles of the type 
$[\xi|x_{2i}\ket{p_2}=0$ are cancelled using similar diagrams with the
leg $p_2$ shifted from the right end to the left end  of the propagator
or vice versa. Eventually however one will run out of free legs to
shift in 
this way and we refer to these situations as boundary cases. The case
where there are no more legs to shift
to the right is covered by the edge diagram as in~\eqref{eq:5}. We
need however to consider the case when there are not enough legs on
the right 
to shift left. More precisely we need to consider the case when there
are just two legs entering the propagator from the right. Then
if we were to shift the final propagator to the left this would leave
a diagram with just 
one leg entering the propagator which vanishes.
So which diagram cancels the spurious pole in this case? The answer is
that the pole is 
canceled by a similar diagram where {\em two} legs have shifted, for example
\begin{align}
\begin{tikzpicture}[x=.7cm,y=.7cm]
%[scale=0.6, every node/.style={scale=0.5}]
  \coordinate[dot,label=below:{}](y2);
  \coordinate[right=2.5 of y2,dotw](y);
  \node[above right=0.3 of y](ang1){};
  \node[left=0.3 of y](ang2){};
  \node[above right=1.5 of y2]{};
  \node[below right=1.5 of y2]{};
  \coordinate[dot,below left=1 of y2](y1);
  \coordinate[dot,above left=1 of y2](y3);
 \node[above right=.7 of y3](p0){};
  \node[above right=1 of y](p1){$p_1$};
%  \node[below=.3 of p2](p3){$p_3$};
  \node[below right=1 of y](p2){$p_2$};
   \node[below right=.7 of y1](pi){};
  \node[below right=1 of y2](p3){$p_3$};
   \coordinate[above left=.2 of y1](y1ext);
  \coordinate[below left=.2 of y3](y3ext);
 \node[below =0.01 of y3ext]{\vdots};
 \begin{pgfonlayer}{background}  \fill[red]
   (y.center)--(ang1.center)--(ang2.center);
 \end{pgfonlayer}
  \draw[very thick,dashed](y2) -- (y);
  \draw[double] (y1ext) -- (y1)--node[auto,inner sep=0]{}
  (y2)--node[auto,inner sep=0]{} (y3)--(y3ext); 
  \draw[photon] (y) -- (p2);
   \draw[photon] (y) -- (p1);
  \draw[photon] (y2) -- (p3);
  \draw[photon] (y1) -- (pi);
  \draw[photon] (y3) -- (p0);
  \node[below right=.6 of p1]{$+$};
  \begin{scope}[xshift=5cm]
      \coordinate[dot,label=below:{}](y2);
  \coordinate[right=2.5 of y2,dotw](y);
  \node[above right=1.5 of y2]{};
  \node[below right=1.5 of y2]{};
  \node[above right=0.3 of y2](ang1){};
  \node[right=0.3 of y2](ang2){};
  \coordinate[dot,below left=1 of y2](y1);
  \coordinate[dot,above left=1 of y2](y3);
  \node[above right=1 of y](p3){$p_2$};
  \node[above right=1 of y2](p2){$p_1$};
  \node[below right=1 of y](pim1){$p_3$};
  \node[above right=.7 of y3](p1){};
  \node[below right=.7 of y1](pi){};
  \coordinate[above left=.2 of y1](y1ext);
  \coordinate[below left=.2 of y3](y3ext);
 \node[below =0.01 of y3ext]{\vdots};
 \begin{pgfonlayer}{background}  \fill[red]
   (y2.center)--(ang1.center)--(ang2.center);
 \end{pgfonlayer}
  \draw[very thick,dashed](y2) -- (y);
  \draw[double] (y1ext) -- (y1)--node[auto,inner sep=0]{}
  (y2)--node[auto,inner sep=0]{} (y3)--(y3ext); 
  \draw[photon] (y) -- (p3);
  \draw[photon] (y2) -- (p2);
  \draw[photon] (y) -- (pim1);
  \draw[photon] (y3) -- (p1);
  \draw[photon] (y1) -- (pi);
  \node[below right=.6 of p3]{$=0$};
  \end{scope}
\end{tikzpicture}
  \end{align}
Here leg $p_1$ has shifted left as expected but leg $p_3$ has
simultaneously shifted right. One can straightforwardly check that the
spurious pole in question cancels between these two diagrams.

Finally then there arises the situation where there is no
leg (such as $p_3$  in the example) to shift to the right in this way. In this final case it turns out
that the spurious pole is canceled by an edge diagram with a single
leg entering the propagator. Such a diagram {\em is} allowed unlike the similar vertex diagram with one leg
entering the propagator.

So for example we get the mutual cancellation of the spurious poles in
the following diagrams:
\begin{align}\label{eq:7}
  \begin{tikzpicture}[>=stealth',x=.7cm,y=.7cm]
          \coordinate[dot,label=below:{}](y2);
  \coordinate[right=2.5 of y2,dotw](y);
  \node[above right=0.3 of y](ang1){};
  \node[left=0.4 of y](ang2){};
  \coordinate[dot,below left=1 of y2](y1);
  \coordinate[dot,above left=1 of y2](y3);
  \node[above right=1 of y](p2){$p_2$};
  \node[above right=1 of y2](p1){$p_1$};
  \node[below right=1 of y](p3){$p_3$};
  \node[above =.5 of y3](p0){};
  \node[below right=.7 of y1](p4){$p_4$};
  \coordinate[above left=.2 of y1](y1ext);
  \coordinate[below left=.2 of y3](y3ext);
 \node[below =0.01 of y3ext]{\vdots};
 \begin{pgfonlayer}{background}  \fill[red]
   (y.center)--(ang1.center)--(ang2.center);
 \end{pgfonlayer}
  \draw[very thick,dashed](y2) -- (y);
  \draw[double] (y1ext) -- (y1)--node[auto,inner sep=0]{}
  (y2)--node[auto,inner sep=0]{} (y3)--(y3ext); 
  \draw[photon] (y) -- (p3);
  \draw[photon] (y2) -- (p1);
  \draw[photon] (y) -- (p2);
  \draw[photon] (y3) -- (p0);
  \draw[photon] (y1) -- (p4);
 \node[below right=.6 of p2]{$+$};
%
%
% PICTURE 2
%
%
  \begin{scope}[xshift=5cm]
    \coordinate[dot,label=below:{}](y2);
  \coordinate[right=2.5 of y2,dotw](y);
   \coordinate[below left=.5 of y2](t);
   \node[above right=0.1 of t](ang1){};
  \node[below left=0.1 of t](ang2){};
  \coordinate (ang3) at ($(t)!0.2!(y)$);
  \coordinate[dot,below left=1 of y2](y1);
  \coordinate[dot,above left=1 of y2](y3);
  \node[right=1 of y](p3){$p_3$};
  \node[above right=1 of y2](p2){$p_2$};
  \node[above =1 of y2](p1){$p_1$};
  \node[below right=.7 of y1](p4){$p_4$};
  \coordinate[above left=.2 of y1](y1ext);
  \coordinate[below left=.2 of y3](y3ext);
 \node[above =.5 of y3](p0){};
  \node[below right=.7 of y1](p4){};
 \node[below =0.01 of y3ext]{\vdots};
 \begin{pgfonlayer}{background}  \fill[red]
   (ang3.center)--(ang1.center)--(ang2.center);
 \end{pgfonlayer}
  \draw[very thick,dashed,<-](t) -- (y);
  \draw[double] (y1ext) -- (y1)--node[auto,inner sep=0]{}
  (y2)--node[auto,inner sep=0]{} (y3)--(y3ext); 
  \draw[photon] (y) -- (p3);
  \draw[photon] (y2) -- (p2);
   \draw[photon] (y2) -- (p1);
  \draw[photon] (y1) -- (p4);
  \draw[photon] (y3) -- (p0);
 \node[right=.2 of p3]{$+$};
  \end{scope}
%
%
%PICTURE 3
%
%
  \begin{scope}[xshift=10cm]
    \coordinate[dot,label=below:{}](y2);
  \coordinate[right=2.5 of y2,dotw](y);
   \coordinate[below left=.5 of y2](t);
  \coordinate[dot,below left=1 of y2](y1);
   \node[below right=0.3 of y](ang1){};
   \coordinate (ang2) at ($(y)!0.2!(y1)$);
  \coordinate[dot,above left=1 of y2](y3);
  \node[above right=1 of y](p3){$p_3$};
 \node[below right=1 of y](p4){$p_4$};
  \node[above right=1 of y2](p2){$p_2$};
  \node[above=1 of y2](p1){$p_1$};
  \node[above=.5 of y3](p0){};
  \coordinate[above left=.2 of y1](y1ext);
  \coordinate[below left=.2 of y3](y3ext);
 \node[below =0.01 of y3ext]{\vdots};
 \begin{pgfonlayer}{background}  \fill[red]
   (y.center)--(ang1.center)--(ang2.center);
 \end{pgfonlayer}
  \draw[very thick,dashed](y1) -- (y);
  \draw[double] (y1ext) -- (y1)--node[auto,inner sep=0]{}
  (y2)--node[auto,inner sep=0]{} (y3)--(y3ext); 
  \draw[photon] (y) -- (p3);
  \draw[photon] (y2) -- (p2);
   \draw[photon] (y2) -- (p1);
  \draw[photon] (y) -- (p4);
  \draw[photon] (y3) -- (p0);
  \node[below right=.6 of p3]{$=0$};
  \end{scope}
  \end{tikzpicture}
\end{align}
If there had been no leg $p_4$ in the above situation, the single
legged propagator would simply continue around the Wilson loop,
cancelling all spurious poles,  until
it did reach a leg.

We note that the edge diagram with
the single leg entering it thus serves a crucial purpose in cancelling
spurious poles, but that these are different types of spurious poles
to the type of pole the generic edge diagrams cancel (compare the
positions of the spurious poles  in~(\ref{eq:7}) with those of~\eqref{eq:5}).

\subsection*{ NMHV$\times$MHV sector}

We also need to consider the various boundary cases of the $W^{(1,0)}$
sector. Here we need to show cancellation of spurious poles at the
level of the integrand. 
Firstly consider the spurious pole cancellation illustrated
in~\eqref{eq:4c}. There we see that spurious poles of the form
$[\xi|q\ket{p_2}=0$ cancel with nearby diagrams where the propagator
slides around the Wilson loop. This procedure continues unimpeded until either
one end of the propagator meets a leg, or the two ends of the
propagator get too close to each other. In the former case there is a
remaining spurious pole with apparently no diagram to cancel it. This  
is precisely the case discussed
in~\eqref{eq:4e}) where one introduces the expression $I$ which cancels
the spurious pole in question. In the latter case we end up with a
propagator stretched between two adjacent vertices. In fact for this
case there is no remaining spurious pole to cancel. This diagram only
has two spurious poles (rather than four) which have already been
cancelled by nearby diagrams. 

So to summarise, in this sector all the boundary cases take care
of themselves, following the introduction of the integrands $I$
discussed in~\eqref{eq:6}.

%\newpage

\end{document}